\documentclass[12pt,preprint]{aastex}
\usepackage{amssymb}
\usepackage{graphicx}

\begin{document}
\title{Hyperaccretion Disks around Neutron Stars}
\author{Dong Zhang and Z. G. Dai}
\affil{Department of Astronomy, Nanjing University, Nanjing 210093,
China; dzg@nju.edu.cn \\ {\rm (Accepted for publication in} The
Astrophysical Journal {\rm August 20, 2008, v683n2 issue)}}

\begin{abstract}
It is usually proposed that hyperaccretion disks surrounding
stellar-mass black holes at an accretion rate of a fraction of one
solar mass per second, which are produced during the mergers of
double compact stars or the collapses of massive stars, are central
engines of gamma-ray bursts (GRBs). In some origin/afterglow models,
however, newborn compact objects are invoked to be neutron stars
rather than black holes. Thus, hyperaccretion disks around neutron
stars seem to exist in some GRBs. Such disks may also occur in
type-II supernovae. In this paper we study the structure of a
hyperaccretion disk around a neutron star. Because of the effect of
a stellar surface, the disk around a neutron star must be different
from that of a black hole. Clearly, far from the neutron star, the
disk may have a flow similar to the black hole disk, if their
accretion rate and central object mass are the same. Near the
compact object, the heat energy in the black-hole disk may be
advected inward to the event horizon, but the heat energy in the
neutron star disk must be eventually released via neutrino emission
because the stellar surface prevents any heat energy from being
advected inward. Accordingly, an energy balance between heating and
cooling would be built in an inner region of the neutron star disk,
which could lead to a self-similar structure of this region. We
therefore consider a steady-state hyperaccretion disk around a
neutron star, and as a reasonable approximation, divide the disk
into two regions, which are called inner and outer disks. The outer
disk is similar to that of a black hole and the inner disk has a
self-similar structure. In order to study physical properties of the
entire disk clearly, we first adopt a simple model, in which some
microphysical processes in the disk are simplified, following Popham
et al. and Narayan et al. Based on these simplifications, we
analytically and numerically investigate the size of the inner disk,
the efficiency of neutrino cooling, and the radial distributions of
the disk density, temperature and pressure. We see that, compared
with the black-hole disk, the neutron star disk can cool more
efficiently and produce a much higher neutrino luminosity. Finally,
we consider an elaborate model with more physical considerations
about the thermodynamics and microphysics in the neutron star disk
(as recently developed in studying the neutrino-cooled disk of a
black hole), and compare this elaborate model with our simple model.
We find that most of the results from these two models are basically
consistent with each other.
\end{abstract}


\section{Introduction}
Gamma-ray bursts (GRBs) are commonly divided into two classes:
short-duration, hard-spectrum bursts, and long-duration,
soft-spectrum bursts. The observations have provided growing
evidence that short bursts result from the mergers of compact star
binaries and long bursts originate from the collapses of massive
stars (for recent reviews see Zhang \& M\'esz\'aros 2004; Piran
2004; M\'esz\'aros 2006; Nakar 2007). It is usually assumed that
both a compact-star merger and a massive-star collapse give rise to
a central black hole and a debris torus around it. The torus has a
mass of about $0.01-1 M_{\odot}$ and a large angular momentum enough
to produce a transient accretion disk with a huge accretion rate up
to $\sim 1.0 M_{\odot}\,{\rm s}^{-1}$. The accretion timescale is
short, e.g., a fraction of one second after the merger of two
neutron stars, and tens of seconds if a disk forms due to fallback
of matter during the collapse process.

The hyperaccretion disk around a black hole is extremely hot and
dense. The optical depth of the accreting gas is so enormous that
radiation is trapped inside the disk and can only be advected
inward. However, in some cases, this hot and dense disk can be
cooled via neutrino emission (Narayan et al. 1992). According to the
disk structure and different cooling mechanisms, flows in the disk
fall into three types: advection-dominated accretion flows (ADAFs),
convection-dominated accretion flows (CDAFs), and neutrino-dominated
accretion flows (NDAFs). The first two types of flow are radiatively
inefficient (Narayan et al. 1998, 2000, 2001) and the final type
cools the disk efficiently via neutrino emission (Popham et al.
1999; Di Matteo et al. 2002). In view of effects of these three
flows, hyperaccretion disks around black holes have been studied
both analytically and numerically (e.g., see Popham et al. 1999;
Narayan et al. 2001; Kohri \& Mineshige 2002; Di Matteo et al. 2002;
Gu et al. 2006; Chen \& Beloborodov 2007; Liu et al. 2007; Janiuk et
al. 2007).

However, newborn neutron stars have been invoked to be central
engines of GRBs in some origin/afterglow models. First, the
discovery of X-ray flares by Swift implies that the central engines
of some GRBs are in a long-living activity (at least hundreds of
seconds) after the bursts (Zhang 2007). This provides a challenge to
conventional hyperaccretion disk models of black holes. Recently,
Dai et al. (2006) argued that newborn central compact objects in the
GRBs could be young neutron stars, at least transiently-existing
neutron stars, rather than black holes (for alternative models see
Perna et al. 2006 and Proga \& Zhang 2006). These neutron stars may
have high angular momentum and their maximum mass may be close to or
slightly larger than the upper mass limit of nonrotating
Tolman-Oppenheimer-Volkoff neutron stars. According to this
argument, Dai et al. (2006) explained X-ray flares of short GRBs as
being due to magnetic reconnection-driven events from
highly-magnetized millisecond pulsars. It is thus reasonable to
assume a unified scenario: a prompt burst originates from a
highly-magnetized, millisecond-period neutron star surrounded by a
transient hyperaccretion disk, and subsequent X-ray flares are due
to a series of magnetic activities of the neutron star.

Second, the shallow decay phase of X-ray afterglows about several
hundreds of seconds after a sizable fraction of GRBs discovered by
Swift has been understood as arising from long-lasting energy
injection to relativistic forward shocks (Zhang 2007). It was
proposed before Swift observations that pulsars in the unified
scenario mentioned above can provide energy injection to a forward
shock through magnetic dipole radiation, leading to flattening of an
afterglow light curve (Dai \& Lu 1998a; Zhang \& M\'esz\'aros 2001;
Dai 2004). Recent model fitting (Fan \& Xu 2006; Yu \& Dai 2007) and
data analysis (Liang et al. 2007) indeed confirm this result. An
ultra-relativistic pulsar wind could be dominated by
electron/positron pairs and its interaction with a postburst
fireball gives rise to a reverse shock and forward shock (Dai 2004).
The high-energy emission due to inverse-Compton scattering in these
shocks is significant enough to be detectable with the upcoming {\em
Gamma-ray Large-Area Space Telescope} (Yu et al. 2007).

Third, we note that, in some origin models of GRBs (e.g., Klu\'zniak
\& Ruderman 1998; Dai \& Lu 1998b; Wheeler et al. 2000; Wang et al.
2000; Paczy\'nski \& Haensel 2005), highly-magnetized neutron stars
or strange quark stars surrounded by hyperaccretion disks resulting
from fallback of matter could occur during the collapses of massive
stars or the mergers of two neutron stars. A similar neutron star
was recently invoked in numerical simulations of Mazzali et al.
(2006) and data analysis of Soderberg et al. (2006) to understand
the properties of supernova SN 2006aj associated with GRB 060218. In
addition, hyperaccretion disks could also occur in type-II
supernovae if fall-back matter has angular momentum. It would be
expected that such disks play an important role in supernova
explosions via neutrino emission, similar to some effects reviewed
by Bethe (1990).

Based on these motivations, we here investigate a hyperaccretion
disk around a neutron star. To our knowledge, this paper is the
first to study hyperaccretion disks around neutron stars related
possibly with GRBs. Chevalier (1996) discussed the structure of
dense and neutrino-cooled disks around neutron stars. He considered
neutrino cooling due to electron-positron pair annihilation, which
is actually much less important than the cooling due to
electron-positron pair capture in the hyperaccreting case of our
interest, since the accretion rate assumed in Chevalier (1996)
($\sim M_{\odot}\,{\rm yr}^{-1}$) is much less than that of our
concern ($\sim 0.1M_{\odot}\,{\rm s}^{-1}$). In this paper we
consider several types of neutrino cooling by using elaborate
formulae developed in recent years. In addition, we focus on some
differences between black hole and neutron star hyperaccretion. A
main difference is that, the internal energy in an accretion flow
may be advected inward into the event horizon without any energy
release if the central object is a black hole, but the internal
energy must be eventually released from the disk if the central
object is a neutron star, because the stellar surface prevents any
heat energy from being advected inward into the star. Since the
accretion rate is always very high, the effective cooling mechanism
in the disk is still neutrino emission, and as a result, the
efficiency of neutrino cooling of the entire disk around a neutron
star should be higher than that of a black hole disk.

There have been a lot of works to study accretion onto neutron stars
in binary systems (e.g., Shapiro \& Salpeter 1975; Klu\'zniak \&
Wilson 1991;  Medvedev \& Narayan 2001; Frank et al. 2002) and
supernova explosions (e.g., Chevalier 1989, 1996; Brown \&
Weingartner 1994; Kohri et al. 2005). In the supernova case,
spherically symmetric accretion onto neutron stars (the so-called
Bondi accretion) was investigated in detail. In particular, Kohri et
al. (2005) tried to use the hyperaccretion disk model with an
outflow wind to explain supernova explosions. In the binary systems,
as the accretion rate is not larger than the Eddington accretion
rate ($\sim 10^{-8}M_{\odot}\,{\rm yr}^{-1}$), the physical
properties of the disk must be very different from those of a
hyperaccretion disk discussed here.

This paper is organized as follows: in \S2, we describe a scheme of
our study of the structure of a quasi-steady disk. We propose that
the disk around a neutron star can be divided into two regions:
inner and outer disks. Table 1 gives the notation and definition of
some quantities in this paper. In order to give clear physical
properties, we first adopt a simple model in \S3, and give both
analytical and numerical results of the disk properties. In \S4, we
study the disk using a state-of-the-art model with lots of elaborate
considerations about the thermodynamics and microphysics, and
compare results from this elaborate model with those from the simple
model. \S5 presents conclusions and discussions.

\section{Description of Our Study Scheme}  
\subsection{Motivations of a two-region disk}  
We study the quasi-steady structure of an accretion disk around a
neutron star with a weak outflow. We take an accretion rate to be a
parameter. For the accretion rates of interest to us, the disk flow
may be an ADAF or NDAF. In this paper, we do not consider a CDAF.
Different from the disk around a black hole, the disk around a
neutron star should eventually release the gravitational binding
energy of accreted matter (which is converted to the internal energy
of the disk and the rotational kinetic energy) more efficiently.

The energy equation of the disk is (Frank et al. 2002)
\begin{equation}
\Sigma v_{r}T\frac{ds}{dr}=Q^{+}-Q^{-},\label{a1}
\end{equation}
where $\Sigma$ is the surface density of the disk, $v_{r}$ is the
radial velocity, $T$ is the temperature, $s$ is the entropy per unit
mass, $r$ is the radius of a certain position in the disk, and
$Q^{+}$ and $Q^{-}$ are the energy input (heating) and the energy
loss (cooling) rate in the disk. From the point of view of
evolution, the structure of a hyperaccretion disk around a neutron
star should be initially similar to that of the disk around a black
hole, because the energy input is mainly due to the local viscous
dissipation, i.e., $Q^{+}=Q^{+}_{\rm vis}$. However, since the
stellar surface prevents the matter and heat energy in the disk from
advection inward any more, a region near the compact object should
be extremely dense and hot as accretion proceeds. Besides the local
viscous heating, this inner region can also be heated by the energy
($Q^{+}_{\rm adv}$) advected from the outer region of the disk.
Thus, the heat energy in this region may include both the energy
generated by itself and the energy advected from the outer region,
that is, we can write $Q^{+}=Q^{+}_{\rm vis}+Q^{+}_{\rm adv}$.
Initially, such a region is so small (i.e., very near the compact
star surface) that it cannot be cooled efficiently for a huge
accretion rate ($\sim 0.01-1M_{\odot}\,{\rm s}^{-1}$). As a result,
it has to expand its size until an energy balance between heating
and cooling is built in this inner region. Such an energy balance
can be expressed by $Q^{+}=Q^{-}$ in the inner region of the disk.

Once this energy balance is built, the disk is in a steady state as
long as the accretion rate is not significantly changed. For such a
steady disk, therefore, the structure of the outer region is still
similar to that of the disk around a black hole, but the inner
region has to be hotter and denser than the disk around a black
hole, and could have a different structure from both its initial
structure and the outer region that is not affected by the neutron
star surface.

Based on the above consideration, the hyperaccretion disk around a
neutron star could have two different regions. In order to discuss
their structure using a mathematical method clearly, as a reasonable
approximation, we here divide the steady accretion disk into an
inner region and an outer region, called inner and outer disks
respectively. The outer disk is similar to that of a black hole. The
inner disk, depending on its heating and cooling mechanisms
discussed above, should satisfy the entropy conservation condition
$Tds/dr\propto Q^{+}-Q^{-}= 0$, and thus we obtain $P\propto
\rho^{\gamma}$, where $P$ and $\rho$ are the pressure and the
density of the disk, and $\gamma$ is the adiabatic index of the disk
gas. Also, the radial momentum equation is
\begin{equation}
(\Omega^{2}-\Omega_{K}^{2})r-\frac{1}{\rho}\frac{d}{dr}(\rho
c_{s}^{2})=0,\label{b01}
\end{equation}
where $\Omega$ and $\Omega_{K}$ are the angular velocity and
Keplerian angular velocity of the inner disk, and
$c_{s}=\sqrt{P/\rho}$ is the isothermal sound speed. We here neglect
the radial velocity term $v_{r}dv_{r}/dr$ since we consider the
situation $v_{r}dv_{r}/dr\ll |\Omega^{2}-\Omega_{K}^{2}|r$, which
can still be satisfied when $v_{r}\ll r\Omega_{K}$ with $\Omega\sim
\Omega_{K}$ but $|\Omega-\Omega_{K}|\geq v_{r}/r$. Equation
(\ref{b01}) gives $\Omega\propto r^{-3/2}$ and $c_{s}\propto
r^{-1/2}$. Moreover, from the continuity equation, we have
$v_{r}\propto (\rho rH)^{-1}$ with the disk's half-thickness
$H=c_{s}/\Omega\propto r$. Thus we can derive a self-similar
structure in the inner region of the hyperaccretion disk of a
neutron star,
\begin{equation}
\rho \propto r^{-1/(\gamma-1)},\,\, P \propto
r^{-\gamma/(\gamma-1)},\,\, v_r \propto
r^{(3-2\gamma)/(\gamma-1)},\label{b02}
\end{equation}
This self-similar structure has been given by Chevalier (1989) and
Brown \& Weingartner (1994) for Bondi accretion under the adiabatic
condition and by Medvedev \& Narayan (2001) and Medvedev (2004) for
disk accretion under the entropy conservation condition. In
addition, if the gas pressure is dominated in the disk, we have
$\gamma=5/3$ so that equation (\ref{b02}) becomes $\rho \propto
r^{-3/2}$, $P \propto r^{-5/2}$, and $v_r \propto r^{-1/2}$, which
have been discussed by Spruit et al. (1987) and Narayan \& Yi
(1994).

An important problem we next solve is to determine the size of the
inner disk. Since the total luminosity of neutrinos emitted from the
whole disk significantly varies with the inner-region size, we can
estimate the inner region size by solving an energy balance between
neutrino cooling and heating in the entire disk. The details will be
discussed in \S 2.3.

Finally, we focus on two problems in this subsection. First, we have
to discuss a physical condition of the boundary layer between the
outer and inner disks. There are two possible boundary conditions.
One condition is to assume that a stalled shock exists at the
boundary layer. Under this assumption, the inner disk is a
post-shock region, and its pressure, temperature and density just
behind the shock are much higher than those of the outer disk in
front of the shock. The other condition is to assume that no shock
exists in the disk, and that all the physical variables of two sides
of this boundary change continuously. We now have to discuss which
condition is reasonable.

Let us assume the mass density, pressure, radial velocity, and
internal energy density of the outer disk along the boundary layer
to be $\rho_{1}$, $P_{1}$, $v_{1}$, and $u_1$, and the corresponding
physical variables of the inner disk to be $\rho_{2}$, $P_{2}$,
$v_{2}$, and $u_2$ at the same radius. Thus the conservation
equations are
\begin{equation}
\begin{array}{ll}\rho_{1}v_{1}=\rho_{2}v_{2}
\\P_{1}+\rho_{1}v_{1}^{2}=P_{2}+\rho_{2}v_{2}^{2}
\\(u_{1}+P_{1}+\rho_{1}v_{1}^{2}/2)/\rho_{1}=(u_{2}+P_{2}
+\rho_{2}v_{2}^{2}/2)/\rho_{2}.
\end{array}
\end{equation}
From the Rankine-Hugoniot relations, we know that if $P_{2}\gg
P_{1}$, the densities of two sides of the boundary layer are
discontinuous, which means that a strong shock exists between the
inner and outer disks. On the other hand, if $P_{1} \sim P_{2}$, we
can obtain $\rho_{1} \sim \rho_{2}$, which means that only a very
weak shock forms at this boundary layer, or we can say that no shock
exists. Therefore, we compare $P_{1}$ and $\rho_{1}v_{1}^{2}$ of the
outer disk. If $P_{1} \ll \rho_{1}v_{1}^{2}$ or $c_{s} \ll v_{1}$,
we can assume that a stalled shock exists at the boundary layer,
i.e., the first boundary condition is correct. If $P_{1} \gg
\rho_{1}v_{1}^{2}$ or $c_{s} \gg v_{1}$, otherwise, we consider
$P_{1} \sim P_{2}$, and thus no shock exists. In $\S$3 and $\S$4, we
will use this method to discuss which boundary condition is
reasonable.

Second, what we want to point out is that the effect of the magnetic
field of the central neutron star is not considered in this paper.
We estimate the order of magnitude of the Alfven radius by using the
following expression (Frank et al 2002), $r_A\simeq
2.07\times10^{4}\dot{m}_{d}^{-2/7}m^{-1/7}\mu_{30}^{4/7}$, where
$\dot{m}_{d}=\dot{M}/0.01M_{\odot}\,{\rm s}^{-1}$ is the accretion
rate, $m=M/1.4M_\odot$ is the mass of the neutron star, and
$\mu_{30}$ is the magnetic moment of the neutron star in units of
$10^{30}{\rm G}\,{\rm cm}^{3}$. Let $r_{*}$ be the neutron star
radius. If the stellar surface magnetic field $B_{s}\leq B_{s,{\rm
cr}}=2.80\times10^{22} \dot{m}_{d}^{1/2}m^{1/4}r_{*}^{-5/4}\,$G or
$B_{s}\leq B_{s,{\rm cr}}=0.89\times10^{15}
\dot{m}_{d}^{1/2}m^{1/4}\,$G when $r_{*}=10^{6}$ cm, then the Alfven
radius $r_A$ is smaller than $r_{*}$. The critical value $B_{s,{\rm
cr}}$ increases with increasing the accretion rate. This implies
that the stellar surface magnetic field affects the structure of the
disk significantly if $B_{s}\geq B_{s,{\rm cr}}\sim
10^{15}-10^{16}\,$G for typical accretion rates. Therefore, we
assume a neutron star with surface magnetic field weaker than
$B_{s,{\rm cr}}$ in this paper. This assumption is consistent with
some GRB-origin models such as Klu\'zniak \& Ruderman (1998), Dai \&
Lu (1998b), Wang et al. (2000), Paczy\'nski \& Haensel (2005), and
Dai et al. (2006), because these models require a neutron star or
strange quark star with surface magnetic field much weaker than
$B_{s,{\rm cr}}$.

\subsection{Structure of the outer disk}   
Here we discuss the structure of the outer disk based on the
Newtonian dynamics and the standard $\alpha$-viscosity disk model
for simplicity. The structure of the hyperaccretion disk around a
stellar-mass black hole has been discussed in many previous works.
The method and equations we use here are similar to those in the
previous works since the outer disk is very similar to the disk
around a black hole.

We approximately consider the angular velocity of the outer disk to
be the Keplerian value $\Omega_K=\sqrt{GM/r^{3}}$. The velocity
$\Omega_K$ should be modified in a relativistic model of accretion
disks (Popham et al. 1999; Chen \& Beloborodov 2007), but we do not
consider it in this paper. We can write four equations to describe
the outer disk, i.e., the continuity equation, the energy equation,
the angular momentum equation and the equation of state.

The continuity equation is
\begin{equation}
\dot{M}=4\pi r\rho v_{r}H\equiv 2\pi r\Sigma v_r,\label{e1}
\end{equation}
where the notations of the physical quantities have been introduced
in $\S$2.1.

In the outer disk, the heat energy could be advected inward and we
take $Q_{\rm adv}^{-}=(1/2)\Sigma Tv_{r}ds/dr$ to be the quantity of
the energy advection rate, where the factor $1/2$ is added because
we only study the vertically-integrated disk over a half-thickness
$H$. Thus the energy-conservation equation (\ref{a1}) is rewritten
as
\begin{equation}
Q^{+}=Q_{\rm rad}^{-}+Q_{\rm adv}^{-}+Q_{\nu}^{-}.\label{e2}
\end{equation}

The quantity $Q^{+}$ in equation (\ref{e2}) is the viscous heat
energy generation rate per unit surface area. According to the
standard viscosity disk model, we have
\begin{equation}
Q^{+}=\frac{3GM\dot{M}}{8\pi r^{3}}f\label{e21}
\end{equation}
where $f=1-(r_{*}/r)^{1/2}$ (Frank et al. 2002).

The quantity $Q_{\rm rad}^{-}$ in equation (\ref{e2}) is the photon
cooling rate per unit surface area of the disk. Since the disk is
extremely dense and hot, the optical depth of photons is always very
large and thus we can take $Q_{\rm rad}^{-}= 0$ as a good
approximation.

The entropy per unit mass of the disk is (similar to Kohri \&
Mineshige 2002)
\begin{equation}
s=s_{\rm gas}+s_{\rm rad}=\frac{S_{\rm gas}}{\rho}+\frac{S_{\rm
rad}}{\rho}=\sum
_{i}n_{i}\left\{\frac{5}{2}\frac{k_{B}}{\rho}+\frac{k_{B}}{\rho}\textrm{ln}\left[\frac{(2\pi
k_{B}T)^{3/2}}{h^{3}n_{i}}\right]\right\}+S_{0}+\frac{2}{3}g_{*}\frac{aT^{3}}{\rho},\label{a01}
\end{equation}
where the summation runs over nucleons and electrons, $k_{B}$ is the
Boltzmann constant, $h$ is the Planck constant, $a$ is the radiation
constant, $S_{0}$ is the integration constant of the gas entropy,
and the term $2g_{*}aT^{3}/3\rho$ is the entropy density of the
radiation with $g_{*}=2$ for photons and $g_{*}=11/2$ for a plasma
of photons and relativistic $e^{+}e^{-}$ pairs. We assume that
electrons and nucleons have the same temperature. Then we use
equation (\ref{a01}) to calculate $ds/dr$ and approximately take
$dT/dr\approx T/r$ and $d\rho/dr\approx \rho/r$ to obtain the energy
advection rate,
\begin{equation}
Q_{\rm
adv}^{-}=v_{r}T\frac{\Sigma}{2r}\left[\frac{R}{2}\left(1+Y_{e}\right)+\frac{4}{3}g_{*}\frac{aT^{3}}{\rho}
\right],\label{a02}
\end{equation}
where the gas constant $R=8.315\times10^{7}$ ergs mole$^{-1}$
K$^{-1}$ and $Y_{e}$ is the ratio of electron to nucleon number
density. The first term in the right-hand bracket of equation
(\ref{a02}) comes from the contribution of the gas entropy and the
second term from the contribution of radiation.

The quantity $Q_{\nu}^{-}$ in equation (\ref{e2}) is the neutrino
cooling rate per unit area. The expression of $Q_{\nu}^{-}$ will be
discussed in detail in $\S$3 and $\S$4.

In this paper we ignore the cooling term of photodisintegration
$Q_{\rm photodis}^{-} $, and approximately take the free nucleon
fraction $X_{\rm nuc}\approx 1$. For the disks formed by the
collapses of massive stars, the photodisintegration process that
breaks down $\alpha$-particles into neutrons and protons is
important in a disk region at very large radius $r$. However, the
effect of photodisintegration becomes less significant for a region
at small radius, where contains less
$\alpha$-particles\footnote{Kohri et al. (2005), Chen \& Beloborodov
(2007), and Liu et al. (2007) discussed the value of nucleon
fraction $X_{\rm nuc}$ and the effect of photodisintegration as a
function of radius for particular parameters such as the accretion
rate and the viscosity parameter $\alpha$. The former two papers
show that $X_{\rm nuc}\approx 1$ and $Q_{\rm photodis}\approx 0$ for
$r\leq 10^{2}r_{g}$. Although the value of $X_{\rm nuc}$ from Liu et
al. (2007) is somewhat different from the previous works, the ratio
of $Q_{\rm photodis}^{-}/Q^{+}$ in their work also drops
dramatically for $r\leq 10^{2}r_{g}$. Therefore, it is convenient
for us to neglect the photodisintegration process for $r\leq
10^{2}r_{g}$ or $r\leq 400$ km for the central star mass
$M=1.4M_{\odot}$.}. On the other hand, for the disks formed by the
mergers of double compact stars, there will be rare
$\alpha$-particles in the entire disk. As a result, we reasonably
take all the nucleons to be free ($X_{\rm nuc}\approx 1$) and
neglect the photodisintegration process, since we mainly focus on
small disks or small regions of the disks (as we will mention in
\S3.4 with the outer boundary $r_{\rm out}$ to be 150 km.)

Furthermore, the angular momentum conservation and the equation of
sate can be written as
\begin{equation}
\nu \Sigma=\frac{\dot{M}}{3\pi}f,\label{e3}
\end{equation}

\begin{equation}
P=P_{e}+P_{\rm nuc}+P_{\rm rad}+P_{\nu},\label{e31}
\end{equation}
where $\nu=\alpha c_{s}H$ in equation (\ref{e3}) is the kinematic
viscosity and $\alpha$ is the classical viscosity parameter, and
$P_{e}$, $P_{\rm nuc}$, $P_{\rm rad}$ and $P_{\nu}$ in equation
(\ref{e31}) are the pressures of electrons, nucleons, radiation and
neutrinos. In $\S$3 we will consider the pressure of electrons in
extreme cases and in $\S$4 we will calculate the $e^{\pm}$ pressure
using the exact Fermi-Dirac distribution function and the condition
of $\beta$-equilibrium.

Equations (\ref{e1}), (\ref{e2}), (\ref{e3}) and (\ref{e31}) are the
basic equations to solve the structure of the outer disk, which is
important for us to study the inner disk.

\subsection{Self-similar structure of the inner disk}  
In \S 2.1 we introduced a self-similar structure of the inner disk,
and described the method to determine the size of the inner disk.
Now we will establish the energy conservation equation in the inner
disk. We assume that $\tilde{r}$ is the radius of the boundary layer
between the inner and outer disks, and that $\tilde {\rho}$, $\tilde
{P}$, and $\tilde {v}_r$ are the density, pressure and radial
velocity of the inner disk just at the boundary layer respectively.
From equation (\ref{b02}), thus, the variables in the inner disk at
any given radius $r$ can be written by
\begin{equation}
\rho=\tilde{\rho}(\tilde{r}/r)^{1/(\gamma-1)}, P=\tilde
{P}(\tilde{r}/r)^{\gamma/(\gamma-1)}, v_r=\tilde{v}_r
(\tilde{r}/r)^{(2\gamma-3)/(\gamma-1)}.\label{en02}
\end{equation}

We take the outer radius of the accretion disk to be $r_{\rm out}$.
The total energy per unit time that the outer disk advects into the
inner disk is (Frank et al. 2002)
\begin{equation}
\dot{E}_{\rm
adv}=(1-\bar{f}_{\nu})\frac{3GM\dot{M}}{4}\left\{{\frac{1}
{\tilde{r}}\left[1-\frac{2}{3}\left(\frac{r_{*}}
{\tilde{r}}\right)^{1/2}\right]-\frac{1}{r_{\rm out}}
\left[1-\frac{2}{3}\left(\frac{r_{*}}{r_{\rm
out}}\right)^{1/2}\right]} \right\},\label{e231}
\end{equation}
where $\bar{f}_{\nu}$ is the average neutrino cooling efficiency of
the outer disk,
\begin{equation}
\bar{f}_{\nu}=\frac{\int_{\tilde{r}}^{r_{\rm out}}Q_{\nu}^{-}2\pi r
dr}{\int_{\tilde{r}}^{r_{\rm out}}Q^{+}2\pi r dr}.\label{ee02}
\end{equation}

If the outer disk flow is mainly an ADAF, $Q_{\nu}^{-}\ll Q^{+}$,
then $\bar{f}_{\nu} \sim 0$; if the outer disk flow is mainly an
NDAF, $Q_{\nu}^{-}\gg Q_{\rm rad}^{-}$ and $Q_{\nu}^{-}\gg Q_{\rm
adv}^{-}$, we have $\bar{f}_{\nu} \sim 1$. The heat energy in the
inner disk should be released more efficiently than the outer disk,
we can still approximately take $\Omega\simeq \Omega_{K}$ in the
inner disk, and the maximum power that the inner disk can release is
\begin{eqnarray}
L_{\nu,max}&\approx&\frac{3GM\dot{M}}{4}\left\{\frac{1}{3r_{*}}-\frac{1}{r_{\rm
out}}\left[1-\frac{2}{3}\left(\frac{r_{*}}{r_{\rm
out}}\right)^{1/2}\right]\right\}\nonumber\\
&&-\bar{f}_{\nu}\frac{3GM\dot{M}}{4}\left\{\frac{1}{\tilde{r}}
\left[1-\frac{2}{3}\left(\frac{r_{*}}{\tilde{r}}\right)^{1/2}\right]-
\frac{1}{r_{\rm out}}\left[1-\frac{2}{3}\left(\frac{r_{*}}{r_{\rm
out}}\right)^{1/2}\right]\right\},
\end{eqnarray}
where we have integrated vertically over the half-thickness. The
first term in the right-hand of this equation is the total heat
energy per unit time of the entire disk, and the second term is the
power taken away through neutrino cooling in the outer disk.
Considering $r_{\rm out}\gg r_*$, we have
\begin{equation}
L_{\nu,max}\approx \frac{3GM\dot{M}}
{4}\left\{\frac{1}{3r_{*}}-\frac{\bar{f}_{\nu}}{\tilde{r}}
\left[1-\frac{2}{3}\left(\frac{r_{*}}{\tilde{r}}\right)^{1/2}\right]\right\}.
\end{equation}

The maximum energy release rate of the inner disk is
$GM\dot{M}/4r_{*}$ if the outer disk flow is mainly an ADAF and
$\bar{f}_{\nu} \sim 0$. This value is just one half of the
gravitational binding energy and satisfies the Virial theorem. If
the outer disk flow is an NDAF, then the energy release of the inner
disk mainly results from the heat energy generated by its own.

Following the above consideration, most of the energy generated in
the disk around a neutron star is still released from the disk, so
we have an energy-conservation equation,
\begin{equation}
\int_{r_{*}}^{\tilde{r}}Q_{\nu}^{-}2\pi rdr=\varepsilon
\frac{3GM\dot{M}}{4} \left\{{\frac{1}{3
r_{*}}-\frac{\bar{f}_\nu}{\tilde{r}}\left[1-\frac{2}{3}
\left(\frac{r_{*}}{\tilde{r}}\right)^{1/2}\right]}\right\},\label{e4}
\end{equation}
where $\varepsilon$ is a parameter that measures the efficiency of
the energy release. If the central compact object is a black hole,
we have $\varepsilon\approx 0$ and the inner disk cannot exist. If
the central compact object is a neutron star, we can take
$\varepsilon\approx 1$ and thus we are able to use equation
(\ref{e4}) to determine the size of the inner disk.

\section{A Simple Model of the Disk} 
In $\S$2, we gave the equations of describing the structure of a
hyperaccretion disk. However, additional equations about
microphysics in the disk are needed. In order to see the physical
properties of the entire disk clearly, we first adopt a simple model
for an analytical purpose. Comparing with \S 4, we here adopt a
relatively simple treatment with the disk microphysics similar to
Popham et al. (1999) and Narayan et al. (2001), and discuss some
important physical properties, and then we compare analytical
results with numerical ones which are also based on the simple
model.

If the disk is optically thin to its own neutrino emission, the
neutrino cooling rate can be written as a summation of four terms
including the electron-positron pair capture rate, the
electron-positron pair annihilation rate, the nucleon bremsstrahlung
rate and the plasmon decay rate, that is, $Q_{\nu}^{-}=(\dot{q}_{\rm
eN}+\dot{q}_{e^{+}e^{-}}+\dot{q}_{\rm brems} +\dot{q}_{\rm
plasmon})H$ (Kohri \& Mineshige 2002). We take two major
contributions of these four terms and use the approximative
formulae:
$\dot{q}_{e^{+}e^{-}}=5\times10^{33}T_{11}^{9}\textrm{ergs\,cm}^{-3}
\,\textrm{s}^{-1}$, and $\dot{q}_{\rm eN}=9\times10^{23}\rho
T_{11}^{6}\textrm{ergs\,cm}^{-3}\,\textrm{s}^{-1}$. Thus equation
({\ref{e2}) can be rewritten as
\begin{equation}
\frac{3GM\dot{M}}{8\pi r^{3}}f=\frac{\dot{M}T}{4\pi
r^{2}}\left[\frac{R}{2}\left(1+Y_{e}\right)+\frac{22}{3}\frac{aT^{3}}{\rho}\right]
+(5\times10^{33}T_{11}^{9}+9\times10^{23}\rho
T_{11}^{6})\frac{c_{s}}{\Omega_K}.\label{e201}
\end{equation}
If neutrinos are trapped in the disk, we use the blackbody limit for
the neutrino emission luminosity: $Q_{\nu}^{-}\sim
(\frac{7}{8}\sigma_{B}T^{4})/\tau$, where $\tau$ is the neutrino
optical depth. We approximately estimate the neutrino optical depth
as $(\dot{q}_{e^{+}e^{-}}+\dot{q}_{\rm
eN})H/(4\times\frac{7}{8}\sigma_{B}T^{4})$.

Moreover, we take the total pressure in the disk to be a summation
of three terms $P_{e}$, $P_{\rm nuc}$ and $P_{\rm rad}$, and neglect
the pressure of neutrinos: $P_{e}=n_{e}kT+K_{1}\left(\rho
Y_{e}\right)^{4/3}$, $P_{\rm nuc}=n_{\rm nuc}kT$, and $P_{\rm
rad}=11aT^{4}/12$, where $K_{1}\left(\rho Y_{e}\right)^{4/3}$ is the
relativistic degeneracy pressure of electrons, $n_e$ and $n_{\rm
nuc}$ are the number densities of electrons and nucleons
respectively. Here we also neglect the non-relativistic degeneracy
pressure of nucleons.
We thus obtain
\begin{equation}
P=P_{e}+P_{\rm nuc}+P_{\rm rad}=\rho
\left(1+Y_{e}\right)RT+K_1\left(\rho Y_{e}
\right)^{4/3}+\frac{11}{12}aT^{4},\label{e32}
\end{equation}
where $K_1=\frac{2\pi hc}{3}\left(\frac{3}{8\pi
m_{p}}\right)^{4/3}=1.24\times10^{15}\,{\rm cgs}$.

Equations (\ref{e3}), (\ref{e201}) and (\ref{e32}) can be solved for
three unknowns (density, temperature and pressure) of the steady
outer disk as functions of radius $r$ for four given parameters
$\alpha$, $Y_{e}$, $\dot{M}$ and $M$ in the simple model. Once the
density, temperature and pressure profiles are determined, we can
present the structure of the outer disk and further establish the
size and the structure of the inner disk.

\subsection{The outer disk}   
We analytically solve equations (\ref{e3}), (\ref{e201}) and
(\ref{e32}) in this subsection. Our method is similar to that of
Narayan et al. (2001). However, what is different from their work is
that we use the same equations to obtain both ADAF and NDAF
solutions in different conditions. Besides, we find that the factor
$f=1-(r_{*}/r)^{1/2}$ cannot be omitted because it plays an
important role in determining the disk structure. In this
subsection, for convenience, we expand the solution range to the
entire disk (i.e., $r_{*}<r<r_{\rm out}$) rather than just consider
it in the outer region. The size of the inner disk, which depends on
the structure of the outer disk, will be solved in $\S$3.2.

First we show a general picture. If the accretion rate is not very
high, most of the energy generated in the disk is advected inward
and we call the disk as an advection-dominated disk. As the
accretion rate increases, the density and temperature of the disk
also increase and the neutrino cooling in some region of the disk
becomes the dominant cooling mechanism. Thus we say that this region
becomes neutrino-dominated. When the accretion rate is sufficiently
large, the disk may totally become neutrino-dominated. Besides the
accretion rate, there are some other factors such as the mass of the
central neutron star, $M$, and the electron-nucleon ratio, $Y_{e}$,
are able to influence the disk structure.

We take the mass of neutron star $M=1.4mM_{\odot}$, the accretion
rate $\dot{M}=\dot{m}_{d}\times 0.01M_{\odot}\,{\rm s}^{-1}$,
$\alpha=0.1\alpha_{-1}$, $r=10^{6}r_{6}\,$cm, $\rho=10^{11}\rho_{11}
\,\textrm{g}\,\textrm{cm}^{-3}$, $T=10^{11}T_{11}\,{\rm K}$, and
$P=10^{29}P_{29}\,\textrm{ergs\,cm}^{-3}$. In the case that the disk
flow is an ADAF with the radiation pressure to be dominated, we find
that the density and temperature in the disk are
\begin{equation}
\begin{array}{ll} \rho_{11}=0.0953 \dot{m}_{d} m^{-1/2}f^{-1/2}\alpha_{-1}^{-1}r_{6}^{-3/2}\\
T_{11}=0.832
m^{1/8}\dot{m}_{d}^{1/4}f^{1/8}\alpha_{-1}^{-1/4}r_{6}^{-5/8}.
\end{array}\label{s11}
\end{equation}
Also, the pressure of the disk from equation (\ref{e32}) becomes
\begin{equation}
P_{29}=3.32m^{1/2} \dot{m}_{d}f^{1/2}\alpha_{-1}^{-1}r_{6}^{-5/2}.
\label{s12}
\end{equation}
We have to check the validity of the assumption made in deriving the
above solution, i.e., we need the relations $Q_{\rm
adv}^{-}>Q_{\nu}^{-}$ and $P_{\rm rad}> P_{\rm gas}$, $P_{\rm rad}>
P_{\rm deg}$ to be satisfied, where $P_{\rm rad}$ and $P_{\rm deg}$
are the gas and degeneracy pressure in the disk. Then we get the
relations that
\begin{equation}
r_{6}f^{1/5}>2.28m^{-3/5}\dot{m}_{d}^{6/5}\alpha_{-1}^{-2}.\label{iq1}
\end{equation}
\begin{equation}
r_{6}f^{-7/3}<74.6\left(1+Y_{e}\right)^{-8/3}m^{7/3}\dot{m}_{d}^{-2/3}\alpha_{-1}^{2/3},\label{iq21}
\end{equation}
\begin{equation}
r_{6}f^{-7/3}<174m^{7/3}\dot{m}_{d}^{-2/3}\alpha_{-1}^{2/3}Y_{e}^{-8/3}.\label{iq22}
\end{equation}
In particular, for a fixed radius $r$, we can rewrite inequation
(\ref{iq1}) as
\begin{equation}
\dot{m}_{d}<0.504m^{1/2}\alpha_{-1}^{5/3}r_{6}^{5/6}f^{1/6}.\label{e301}
\end{equation}
which means that a larger radius allows a larger upper limit of the
accretion rate for the radiation-pressure-dominated region. On the
other hand, if the parameters $m$, $\alpha$ and $\dot{m}_{d}$ in
some region of the disk do not satisfy inequations (\ref{iq1}) or
(\ref{e301}), the other types of pressure can exceed the radiation
pressure but the region can still be advection-dominated. For an
analytical purpose, we want to discuss the range of different types
of pressure in two extreme cases where $Y_{e}\sim 1$ or $Y_{e}\ll
1$. From inequation (\ref{iq21}), we can see that, since the minimum
value of $r_{6}f^{-7/3}$ is 19.9, when $\dot{m}_{d}>
0.453m^{7/2}\alpha_{-1}$ for $Y_{e}\sim 1$, or $\dot{m}_{d}>
7.25m^{7/2}\alpha_{-1}$ for $Y_{e}\ll 1$ , the gas pressure takes
over the radiation pressure in the disk and the entire disk becomes
gas-pressure-dominated. On the other hand, the degeneracy pressure
is larger than the radiation pressure at a very large radius if the
electron fraction $Y_{e}$ is not very small. However, we do not
consider this situation for an ADAF region of the disk, because the
degeneracy pressure, even if larger than the radiation pressure,
cannot exceed the gas pressure.

When the gas pressure is dominated and the outer disk flow is still
an ADAF, we obtain the density, temperature and pressure of the disk
as
\begin{equation}
\begin{array}{ll} \rho_{11}=0.556\left(1+Y_{e}\right)^{-12/11}
m^{5/11}\dot{m}_{d}^{8/11}f^{5/11}\alpha_{-1}^{-8/11}r_{6}^{-21/11}\\
T_{11}=1.29\left(1+Y_{e}\right)^{-3/11}m^{4/11}
\dot{m}_{d}^{2/11}f^{4/11}\alpha_{-1}^{-2/11}r_{6}^{-8/11}\\
P_{29}=5.98\left(1+Y_{e}\right)^{-4/11}m^{9/11}
\dot{m}_{d}^{10/11}f^{9/11}\alpha_{-1}^{-10/11}r_{6}^{-29/11} .
\end{array}\label{e302}
\end{equation}
Similarly, we discuss the validity of the solution (\ref{e302}). The
assumption of a gas-pressure-dominated ADAF disk ($Q^{-}_{\rm
adv}>Q^{-}_{\nu}$) requires
\begin{equation}
r_{6}^{47/22}f^{-20/11}>128\left(1+Y_{e}
\right)^{-26/11}m^{29/22}\dot{m}_{d}^{10/11}\alpha_{-1}^{-21/11}.\label{e303}
\end{equation}
$P_{\rm gas}>P_{\rm rad}$ can be written by
\begin{equation}
r_{6}f^{-7/3}>74.6\left(1+Y_{e}\right)^{-8/3}
m^{7/3}\dot{m}_{d}^{-2/3}\alpha_{-1}^{2/3}.\label{iq22}
\end{equation}
$P_{\rm gas}>P_{\rm deg}$ leads to
\begin{equation}
r_{6}f^{-7/3}<8.07\times10^{3}\left(1+Y_{e}\right)^{12}Y_{e}^{-44/3}m^{7/3}\dot{m}_{d}^{-2/3}\alpha_{-1}^{2/3},\label{e304}
\end{equation}
which is satisfied for a large parameter space.

If $Y_{e}\sim 1$ and the parameters $m$ and $\alpha$ make
$0.453m^{7/2}\alpha_{-1}<\dot{m}_{d}<1.73m^{-29/20}\alpha_{-1}^{21/10}$
valid, the entire disk becomes an advection-dominated disk with the
gas pressure to be dominated. However, if $Y_{e}\ll 1$, such a disk
cannot exist, since inequation (\ref{e303}) cannot be always
satisfied in the entire disk and some region of the disk would
become neutrino-dominated. Also, when the mass accretion rate
$\dot{m}_{d}$ becomes higher, most region of the disk becomes
neutrino-dominated.

Furthermore, in the region where neutrino cooling is efficient and
the gas pressure dominates over the degeneracy pressure, we have
another particular solution
\begin{equation}
\begin{array}{ll} \rho_{11}=2.38\left(1+Y_{e}\right)^{-9/5}m^{17/20}\dot{m}_{d}f\alpha_{-1}^{-13/10}r_{6}^{-51/20} \\
T_{11}=0.490\left(1+Y_{e}\right)^{1/5}m^{1/10}
\alpha_{-1}^{1/5}r_{6}^{-3/10} \\
P_{29}=9.71\left(1+Y_{e}\right)^{-3/5}m^{19/20}
\dot{m}_{d}f\alpha_{-1}^{-11/10}r_{6}^{-57/20}.
\end{array}\label{s2}
\end{equation}
In this case, the temperature is independent of the accretion rate
$\dot{m}_{d}$. Finally we check the gas pressure-dominated
assumption. Using equation (\ref{s2}) and assuming $P_{\rm
gas}>P_{\rm deg}$, we obtain
\begin{equation}
r_{6}f^{-20/33}>3.18Y_{e}^{80/33}\left(1+Y_{e}
\right)^{-36/11}m^{1/3}\dot{m}_{d}^{20/33}\alpha_{-1}^{-38/33}.
\label{e5}
\end{equation}
Inequation (\ref{e5}) is always satisfied if $Y_{e}\ll 1$, and thus
we can say that the gas pressure-dominated assumption is valid.
However, if $Y_{e}\sim 1$, a part of the disk becomes degeneracy
pressure-dominated if
$\dot{m}_{d}>64.5\alpha_{-1}^{19/10}m^{-11/20}$. In particular, in
the case of $Y_{e}\sim 1$ and large accretion rate $\dot{m}_{d}$, a
set of solutions on the part of the disk are
\begin{equation}
\begin{array}{ll} \rho_{11}=1.26Y_{e}^{-4/3}m^{2/3}\dot{m}_{d}^{2/3}
f^{2/3}\alpha_{-1}^{-2/3}r_{6}^{-2} \\
T_{11}=0.526Y_{e}^{4/27}m^{13/108}\dot{m}_{d}^{1/27}f^{1/27}\alpha_{-1}^{7/54}
r_{6}^{-13/36} \\
P_{29}=7.85Y_{e}^{-4/9}m^{8/9}\dot{m}_{d}^{8/9}f^{8/9}
\alpha_{-1}^{-8/9}r_{6}^{-8/3}.\label{e3031}
\end{array}
\end{equation}
which describe an NDAF with the degeneracy pressure to be dominated.
However, we should remember that in deriving the above solutions we
have not considered the constraint of $r>\tilde{r}$.

Here we make a summary of $\S$ 3.1. We used an analytical method to
solve the density, pressure and temperature of the outer disk based
on a simple model discussed at the beginning of $\S$3. The accretion
flow may be ADAF or NDAF with the radiation, gas or degeneracy
pressure to be dominated. We fix radius $r$ and show several
possible cases in the disk with different accretion rate
$\dot{m}_{d}$ in Table 2. Moreover, for different electron fraction
$Y_{e}$ and fixing $m=1$ and $\alpha_{-1}=1$, we calculate the upper
limit of $\dot{m}_{d}$. If $Y_{e}\sim 1$, the advection-dominated
region in the disk can be radiation or gas pressure-dominated, and
the neutrino-cooled region can be gas or degeneracy
pressure-dominated. However, if $Y_{e}\ll 1$, the gas
pressure-dominated region in the ADAF case is very small, and the
degeneracy pressure-dominated region cannot exist. In \S 3.4 we will
obtain similar results by using a numerical method.

The solutions given in this subsection can also be used to discuss
the properties of the disk around a black hole. Our analytical
solutions of the outer disk are similar to those of Narayan et al.
(2001) and Di Matteo et al. (2002), who took $Y_{e}$ to be a
parameter. Narayan et al. (2001) found that advection-dominated
disks can be radiation or gas pressure-dominated, and
neutrino-dominated disks can be gas or degeneracy pressure-dominated
instead. This is consistent with our conclusion for $Y_{e} \sim 1$.
However, these authors did not consider the factor
$f=1-\sqrt{r_{*}/r}$, which is an important factor because a small
$r$, as we have mentioned above, can dramatically change the
parameter space of the outer disk. Di Matteo et al. (2002) discussed
different pressure components (their Fig. 2), which is also
consistent with our conclusion. Chen \& Beloborodov (2007)
calculated the value of $Y_{e}$ and showed that $Y_{e}\ll 1$ when
$r$ is small. According to the above discussion, therefore, the
degeneracy-pressure-dominated region in the NDAF disk cannot exist.
This is consistent with Chen \& Beloborodov (2007) that the pressure
in a neutrino-cooled disk is dominated by baryons (gas).

However, our analytical results are partly different with Kohri et
al. (2005) and Chen \& Beloborodov (2007) which showed that the
electron pressure is dominant in some advection-dominated regions of
the disk. This difference is mainly because that we take $P_{\rm
rad}=11aT^{4}/12$ in our analytical model, which includes the
contribution of relativistic electron-positron pairs. However, Kohri
et al. (2005) and Chen \& Beloborodov (2007) took $P_{\rm
rad}=aT^{4}/3$ and calculated the pressure of $e^{+}e^{-}$ pairs in
the term of electron pressure $P_{e}$. As a result, the radiation
pressure we consider in this subsection is actually the pressure of
a photon and $e^{+}e^{-}$-pair plasma.

In the following subsection, we will establish the structure of the
inner disk depending on the outer disk solutions here, and use the
value of $\tilde{r}$ to further constrain the solutions that we have
obtained.

\subsection{The inner disk}  
\subsubsection{Boundary layer between the inner disk and outer disk}
We use the method discussed in \S 2.1 and compare the radial
velocity with the local speed of sound of the outer disk using the
results given in section \S 3.1.

The radial velocity of the outer disk at radius $r$ is
\begin{equation}
v_{1}=\frac{\dot{M}}{2\pi r\Sigma}=\frac{\dot{M}}{4\pi r\rho_{1}
H}=\frac{\dot{M}\sqrt{GM}}{4\pi r^{5/2}\sqrt{P_{1} \rho_{1}}}.
\end{equation}
Hence, we have
\begin{equation}
\frac{v_{1}^{2}}{c_{s}^{2}}\sim
\frac{\rho_{1}v_{1}^{2}}{P_{1}}=\frac{\dot{M}^{2}GM}{16\pi^{2}r^{5}P_{1}^{2}}
=0.0465\frac{\dot{m}_{d}^{2}m}{r_{6}^{5}P_{1,29}^{2}}.
\end{equation}

In the case where the outer disk is ADAF and the radiation pressure
is dominated, using solutions (\ref{s11}) and (\ref{s12}), we have
\begin{equation}
\frac{v_{1}^{2}}{c_{s}^{2}}\sim
4.22\times10^{-3}\alpha_{-1}^{2}f^{-1}.
\end{equation}
Therefore, we see $v_{1}\ll c_{s}$ for typical values of the
parameters.

In the case of ADAF with the gas pressure to be dominated, using
solution (\ref{e302}), we have
\begin{equation}
\frac{v_{1}^{2}}{c_{s}^{2}}\sim
1.302\times10^{-3}\left(1+Y_{e}\right)^{8/11}
m^{-7/11}\dot{m}_{d}^{2/11}\alpha_{-1}^{20/11}f^{-18/11}r_{6}^{3/11}.\label{e3211}
\end{equation}

For NDAF, using expression (\ref{s2}) to compare the radial velocity
with the speed of sound, we have
\begin{equation}
\frac{v_{1}^{2}}{c_{s}^{2}}\sim
4.94\times10^{-4}\left(1+Y_{e}\right)^{6/5}
m^{-9/10}\alpha_{-1}^{11/5}f^{-2}r_{6}^{7/10}.\label{e3212}
\end{equation}
Note that this ratio is independent of the accretion rate. From
equations (\ref{e3211}) and (\ref{e3212}), we still find that
$v_{1}\ll c_{s}$ is always satisfied except for a region very near
the stellar surface, where the factor $f$ is very small. This
region, however, is so small that it belongs to the inner disk where
we have to use the self-similar structure, which we will discuss
later in details.

Therefore, for the hyperaccretion disk discussed in this paper, as
the disk is extremely hot and dense, the radial velocity is always
subsonic. So there is no stalled shock between the inner and outer
disks. Thus, all physical variables between two sides of the
boundary layer between two regions of the disk change continuously.
Besides, the rotational velocity is assumed to be the Keplerian
value and has no jump at the boundary layer.

\subsubsection{Solution of the inner disk }  
We now study the inner disk analytically based on the results given
in $\S$3.1. The main problem that we should solve in this subsection
is to determine the size of the inner disk for a range of parameters
and to describe the structure of the inner disk. In the case where
the radiation pressure is dominated, by using the self-similar
structure (\ref{en02}), we obtain the temperature of the inner disk,
\begin{equation}
T_{11}=\left(\frac{P_{29}}{6.931}\right)^{1/4}=1.01m^{1/6}\tilde{\rho}_{11}^{1/2}\dot{m}_{d}^{1/6}
\tilde{f}^{1/6}\alpha_{-1}^{-1/6}\tilde{r}_{6}^{\frac{2-\gamma}{4(\gamma-1)}}r_{6}^{\frac{-\gamma}{4(\gamma-1)}}.\label{e11}
\end{equation}
where $\tilde{r}$ and $\tilde{\rho}$ are the radius and density of
the boundary layer between the inner and outer disk, and
$\tilde{f}=1-(r_{*}/\tilde{r})^{1/2}$.

Using the self-similar condition and the above expression of
$T_{11}$, we find the total neutrino cooling rate,
\begin{equation}
\int_{r_{*}}^{\tilde{r}}Q_{\nu}^{-}2\pi
rdr=10^{51}\times\int120\tilde{\rho}_{11}^{7/6}
m^{5/6}\dot{m}_{d}^{4/3}\alpha_{-1}^{-4/3}\tilde{f}^{4/3}
\tilde{r}_{6}^{\frac{9-4\gamma}{2(\gamma-1)}}r_{6}^{\frac{\gamma-6}{2(\gamma-1)}}dr_{6}.
\end{equation}

The outer disk flow is mainly an ADAF, using the solution of a
radiation pressure-dominated ADAF (i.e., solutions \ref{s11} and
\ref{s12}), we have the total neutrino cooling rate,
\begin{eqnarray}
L_\nu  = 1.55\times10^{52}\,{\rm ergs}\,{\rm
s}^{-1}\left(\frac{\gamma-1}{8-3\gamma}\right)
m^{1/4}\dot{m}_{d}^{5/2}
 \tilde{f}^{3/4} \alpha_{-1}^{-5/2}\tilde{r}_{6}^{\frac{25-15\gamma}
{4(\gamma-1)}}\left(r_{*,6}^{\frac{3\gamma-8}{2(\gamma-1)}}-
\tilde{r}_{6}^{\frac{3\gamma-8}{2(\gamma-1)}}\right).
\end{eqnarray}
From the energy-conservation equation (\ref{e4}), the position of
the boundary layer satisfies the following equation,
\begin{equation}
\tilde{r}_{6}^{\frac{5(5-3\gamma)}{4(\gamma-1)}}\left(1-\sqrt{\frac{r_{*,6}}{\tilde{r}_{6}}}\right)^{3/4}
\left(r_{*,6}^{\frac{3\gamma-8}{2(\gamma-1)}}-
\tilde{r}_{6}^{\frac{3\gamma-8}{2(\gamma-1)}}\right)
=0.0597\left(\frac{8-3\gamma}{\gamma-1}\right)m^{3/4}\dot{m}_{d}^{-3/2}\alpha_{-1}^{5/2}r_{*,6}^{-1},\label{e6}
\end{equation}
where we take $\varepsilon\sim 1$ and $\bar{f}_{\nu}\sim 0$ in
equation (\ref{e4}). The left-hand term of equation (\ref{e6})
increases with increasing $\tilde{r}_{6}$, so $\tilde{r}_{6}$
decreases if $\dot{m_{d}}$ increases in the right-hand term of this
equation. In other words, the size of the inner disk decreases as
the accretion rate increases. From equation (\ref{e6}), we can also
see that its solution, $\tilde{r}_{6}$, is independent of $Y_{e}$,
and increases with the mass of the central star. In addition, since
the gas pressure has its own contribution to the disk, the actual
adiabatic index $\gamma$ is larger than 4/3, which can makes the
solution of equation (\ref{e6}) larger. For an analytical purpose,
we assume several different sets of parameters to solve equation
(\ref{e6}). Table 3 clearly shows that, in the
radiation-pressure-dominated disk with an advection-dominated outer
region, as the accretion rate increases, the value of $\tilde{r}$
decreases, and that $\tilde{r}$ increases with increasing $\gamma$
or decreasing $m$.

In the case where the gas pressure is dominated and the outer disk
flow is still an ADAF, we obtain the temperature of the inner disk,
\begin{equation}
T_{11}=\left(1+Y_{e}\right)^{-1}\frac{1}{8.315}
\frac{P_{29}}{\rho_{11}} =\left(1+Y_{e}\right)^{-1}0.874m^{2/3}
\tilde{\rho}_{11}^{-2/3}\dot{m}_{d}^{2/3}\tilde{f}^{2/3}
\alpha_{-1}^{-2/3}\tilde{r}_{6}^{-1}r_{6}^{-1}.
\end{equation}
Similarly, from equations (\ref{e4}) and (\ref{e302}), we have
\begin{equation}
\int_{r_{*}}^{\tilde{r}}Q_{\nu}^{-}2\pi
rdr=10^{51}\times\int45.0\tilde{\rho}_{11}^{-10/3}\left(1+Y_{e}\right)^{-6}m^{23/6}\dot{m}_{d}^{13/3}\tilde{f}^{13/3}
\tilde{r}_{6}^{(\frac{1}{\gamma-1}-\frac{13}{2})}r_{6}^{-(4+\frac{1}
{\gamma-1})}dr_{6}.
\end{equation}
The neutrino luminosity of the inner disk reads
\begin{eqnarray}
L_\nu & = & 3.53\times10^{53}\,{\rm ergs}\,{\rm
s}^{-1}\left(\frac{\gamma-1}{3\gamma-2}\right)
\left(1+Y_{e}\right)^{-26/11}\nonumber \\ & & \times
m^{51/22}\dot{m}_{d}^{21/11}
\tilde{f}^{\frac{31}{11}}\alpha_{-1}^{-21/11}\tilde{r}_{6}^{(\frac{1}
{\gamma-1}-\frac{3}{22})}\left(r_{*,6}^{\frac{2-3\gamma}{\gamma-1}}-
\tilde{r}_{6}^{\frac{2-3\gamma}{\gamma-1}}\right).
\end{eqnarray}
The energy-conservation equation of the inner disk is
\begin{eqnarray}
\left(1-\sqrt{\frac{r_{*,6}}{\tilde{r}_{6}}}\right)^{-31/11}\tilde{r}_{6}^{(\frac{-1}
{\gamma-1}+\frac{3}{22})}\left(r_{*,6}^{\frac{2-3\gamma}{\gamma-1}}-
\tilde{r}_{6}^{\frac{2-3\gamma}{\gamma-1}}\right)^{-1}
\nonumber\\
=\frac{382(\gamma-1)}{3\gamma-2}\left(1+Y_{e}\right)^{-26/11}
m^{29/22}\dot{m}_{d}^{10/11}\alpha_{-1}^{-21/11}r_{*,6}.\label{e321}
\end{eqnarray}

From equation (\ref{e321}), we see that the size of the inner disk
($\tilde{r}$) also decreases with increasing the accretion rate
$\dot{m}_{d}$. Table 4 gives solutions of equation (\ref{e321}) with
different sets of parameters. We can see that $\tilde{r}$ also
decrease with increasing $\tilde{\gamma}$, $m$ or decreasing
$Y_{e}$.

We above study the case where the outer disk is advection-dominated,
and find that the size of the inner disk always increase with the
accretion rate. In the case where the outer disk is mainly
neutrino-dominated, using expression (\ref{s2}) and equation
(\ref{e4}), we obtain an energy-conservation equation in the inner
disk,
\begin{equation}
2.77\left(\frac{\gamma-1}{3\gamma-2}\right)
\tilde{r}_{6}^{\frac{2\gamma-1}{\gamma-1}}
\tilde{f}\left(r_{*,6}^{\frac{2-3\gamma}{\gamma-1}}-
\tilde{r}_{6}^{\frac{2-3\gamma}{\gamma-1}}\right)
=0.924\left\{\frac{1}{r_{*,6}}-\frac{3\bar{f}_{\nu}}{\tilde{r}_{6}}\left[1-\frac{2}{3}
\left(\frac{r_{*,6}}{\tilde{r}_{6}}\right)^{1/2}\right]\right\}.\label{e7}
\end{equation}

This equation shows us that $\tilde{r}_{6}$ in NDAF is independent
of $Y_{e}$, $m$, $\dot{m}_{d}$, and $\alpha$, but only dependent on
$\gamma$ and $\bar{f}_{\nu}$. The size of the inner disk is constant
no matter how much the components, the accretion rate of the disk,
and the mass of the central neutron star are. If the outer disk is
mainly an NDAF, we have $\bar{f}_{\nu}\sim 1$. We choose several
different sets of parameters to obtain the solution of equation
(\ref{e7}) (see Table 5). As $\bar{f}_{\nu}$ increases, the value of
$\tilde{r}$ decreases slightly. Here we also consider an
intermediate case of $\gamma$ between $5/3$ and $4/3$, the decline
of $\gamma$ makes the size of the inner disk decrease.

However, in the above discussion about an NDAF, we have not
considered the effect of neutrino opacity but simply assumed that
neutrinos escape freely. Actually, if the accretion rate is
sufficiently large and the disk flow is mainly an NDAF, and the
disk's region near the neutron star surface can be optically thick
to neutrino emission. With increasing the accretion rate, the area
of this optically thick region increases. We now estimate the effect
of neutrino opacity on the structure of the inner disk. Let the
region of $r_{*}<r<\bar{r}$ be optically thick to neutrino emission,
the region of $\bar{r}<r<\tilde{r}$ be optically thin, and the
electron/positron pair capture reactions be the dominant cooling
mechanism. Thus equation (\ref{e4}) becomes
\begin{equation}
\int_{r_{*}}^{\bar{r}}\frac{4\left(\frac{7}{8}\sigma_{B}T^{4}\right)}{\tau}2\pi
rdr+\int_{\bar{r}}^{\tilde{r}}9\times10^{34}\rho_{11}T_{11}^{6}H2
\pi rdr =\frac{3GM\dot{M}}{4} \left\{{\frac{1}{3
r_{*}}-\frac{\bar{f}_{\nu}}{\tilde{r}}\left[1-\frac{2}{3}\left(\frac{r_{*}}
{\tilde{r}}\right)^{1/2}\right]}\right\}.\label{e10}
\end{equation}
where we take $\varepsilon\approx 1$. Using the self-similar
relations and performing some derivations, we find
\begin{eqnarray}
18.7\left(1+Y_{e}\right)^{8/5}\left(\frac{\gamma-1}{2-\gamma}\right)
m^{-1/5}\dot{m}_{d}^{-1}\tilde{f}^{-1}\alpha_{-1}^{8/5}
\tilde{r}_{6}^{(\frac{18}{5}-\frac{1}{\gamma-1})}
\left[\bar{r}_{6}^{-1+1/(\gamma-1)}
-r_{*,6}^{-1+1/(\gamma-1)}\right]
\nonumber\\+2.77\left(\frac{\gamma-1}{3\gamma-2}\right)m\dot{m}_{d}\tilde{f}
\tilde{r}_{6}^{\frac{2\gamma-1}{\gamma-1}}\left(\bar{r}_{6}^{\frac{2-3\gamma}
{\gamma-1}}-\tilde{r}_{6}^{\frac{2-3\gamma}{\gamma-1}}\right)
=0.924m\dot{m}_{d}\left[\frac{1}{r_{*,6}}-\frac{3\bar{f}_{\nu}}
{\tilde{r}_{6}}+\frac{2\bar{f}_{\nu}}{\tilde{r}_{6}}\left(\frac{r_{*,6}}
{\tilde{r}_{6}}\right)^{1/2}\right]. \label{e8}
\end{eqnarray}
The solution shown by equation (\ref{e8}) gives $\tilde{r}$. We take
$\bar{f}_{\nu}\approx 1$ and $\alpha=0.1$. Moreover, we define a new
parameter $k=\bar{r}_{6}/\tilde{r}_{6}$, and assume several sets of
parameters to give the solution of equation (\ref{e8}).

From Table 6, we can see that the size of the inner disk increases
with the accretion rate. In addition, an increase of $m$ or $k$, or
an decrease of $Y_{e}$ also makes the inner-disk size larger. We
will compare the analytical results from Tables 3 to 6 with
numerical results in $\S$3.4 in more details.

\subsection{Discussion about the stellar surface boundary}  
Now we want to discuss the physical condition near the neutron star
surface. We know that if the rotational velocity of the neutron star
surface is different from that of the inner boundary of the disk,
then the disk can act a torque on the star at the stellar radius.
Here we take the rotational velocity of the inner disk $\Omega\simeq
\Omega_{K}$ approximately as mentioned in \S 2.1 and \S 2.3. Then we
always have the stellar surface angular velocity $\Omega_{*}$ to be
slower than that of the inner disk $\Omega_{K}$ (i.e., $\Omega_{*}<
\Omega_{K}$). As a result, the kinetic energy of the accreted matter
is released when the angular velocity of the matter decreases to the
angular velocity of the neutron star surface. From the Newtonian
dynamics, we obtain a differential equation,
\begin{equation}
G_{*}r_{*}=\frac{d(I\Omega_{*})}{dt},\label{e331}
\end{equation}
where $G_{*}r_{*}=\dot{M}r_{*}^{2}(\Omega_K-\Omega_{*})$ is the
torque acting on the star surface from the disk, $I$ is the moment
of inertia of the star, $I=\xi M r_{*}^{2}$, with $\xi$ being a
coefficient. The above equation can be further written as
\begin{equation}
[\Omega_K-\Omega_{*}(1+\xi)]dM=\xi Md\Omega_{*}. \label{s3}
\end{equation}
From equation (\ref{s3}) we can see that if
$\Omega_{*,0}<\Omega_K/(1+\xi)$ initially, we always have
$d\Omega_{*}/dM>0$, i.e., the neutron star is spun up as accretion
proceeds. On the other hand, if $\Omega_{*,0}>\Omega_K/(1+\xi)$
initially, then the star is always spun down by the disk. The limit
value of $\Omega_{*}$ is $\Omega_K/(1+\xi)$ in both cases. The
solution of equation (\ref{s3}) is
\begin{equation}
\left|\Omega_{*}-\frac{\Omega_K}{1+\xi}\right|=\left(\frac{M_{0}}{M_{0}+\Delta
M}\right)^{\frac{\xi+1}{\xi}}\left|\Omega_{*,0}-\frac{\Omega_K}{1+\xi}\right|\approx
\left[1-\left(\frac{\xi+1}{\xi}\right)\frac{\Delta
M}{M_{0}}\right]\left|\Omega_{*,0}-\frac{\Omega_K}{1+\xi}\right|,\label{e9}
\end{equation}
where $M_{0}$ is the initial mass of the neutron star, and $\Delta
M$ is the mass of the accreted matter from the disk by the star. The
change of $\Omega_{*}$ depends on the ratio $\Delta M/M_{0}$. For
example, if we assume $M_{0}=1.4M_{\odot}$, $\Delta M=0.01M_{\odot}$
and set $\xi=2/5$, then we have
$(\frac{\Omega_K}{1+\xi}-\Omega_{*})=0.98(\frac{\Omega_K}{1+\xi}-\Omega_{*,0})$.
Or if we take $\Delta M=0.1M_{\odot}$, $\xi=2/5$, and the central
stellar mass is unchanged, then we obtain
$(\frac{\Omega_K}{1+\xi}-\Omega_{*})=0.79(\frac{\Omega_K}{1+\xi}-\Omega_{*,0})$.

Here we consider that the energy released from the surface boundary
is also taken away by neutrino emission, and assume that neutrinos
emitted around the stellar surface are opaque. We can estimate the
temperature of the neutron star surface through
$\frac{7}{8}\sigma_{B}T^{4}4\pi r_{*}H\sim
\frac{GM\dot{M}}{2r_{*}}(2-\varepsilon)$, where $H$ is the half
thickness of the inner boundary of the disk, and the parameter
$\varepsilon$ has the same meaning as that in $\S$2.3. Then we can
estimate the surface temperature as $T_{11}\sim
0.415(2-\varepsilon)^{1/4}m^{1/4}\dot{m}_{d}^{1/4}H_{6}^{-1/4}r_{*,6}^{-1/2}$.
This estimation of the temperature is only valid if the inner disk
is optically thin for neutrinos and only the surface boundary is
optically thick. If the inner disk, due to a large accretion rate,
becomes optically thick to neutrino emission, the surface
temperature should be higher.

\subsection{Comparison with numerical results}
In order to give an analytical solution of the accretion disk around
a neutron star in the simple model, we can choose the dominant terms
in equations (\ref{e201}) and (\ref{e32}). We now consider one type
of pressure to be dominated, and assume extreme cases from ADAF to
NDAF. For example, we consider the neutrino-dominated region with
$\bar{f}_{\nu}\sim1$ and the advection-dominated region with
$\bar{f}_{\nu}=0$. In this subsection, we solve the
hyperaccretion-disk structure numerically based on the simple model.
In order to compare the numerical results with what we have obtained
analytically, we keep on with using the equations and definitions of
all the parameters in this section. However, we first need to point
out several approximations and some differences between numerical
and analytical methods based on the simple model.

First of all, we choose the range of $r_{6}$ from 1 to 15 in our
numerical calculations. In other words, we take the range of the
accretion disk to be from the surface of the neutron star to the
radius of 150 km as the outer boundary. During the compact-star
merger or massive-star collapse, the torus around a neutron star has
only some part that owns a large angular momentum to form a debris
disk, so the mass of the disk may be smaller than the total mass of
the torus. From view of simulations (Lee \& Ramirez-Ruiz 2007), if
the debris disk forms through the merger of two neutron stars, its
outer radius can be slightly smaller than the value we give above.
On the other hand, the disk size may be slightly larger than that we
assume above if the disk forms during the collapse of a massive
star. The changes of physical variables of the disk due to a change
of the outer radius may be insignificant, and we do not discuss the
effect of the outer radius in this paper.

Second, we fix the viscosity parameter $\alpha$ to be 0.1. If
$\alpha$ decreases (or increases), the variables of the disk
increases (or decreases). More information can be seen in analytical
solutions in \S 3.1 and \S 3.2. In numerical calculations, we take
$\alpha$ to be a constant.

Third, we still set $Y_{e}$ as a parameter in numerical calculations
in \S 3.4. In order to show results clearly, we consider two
conditions: one is an extreme condition with $Y_{e}=1$, which means
that the disk is made mainly of electrons and protons but no
neutrons; the other is $Y_{e}=1/9$, which means that the number
ratio of electrons, protons and neutrons is $1:1:8$. An elaborate
work should consider the effect of $\beta$-equilibrium, and we will
discuss it in detail in $\S$4.

For numerical calculations, we consider all the terms of the
pressure and an intermediate case between ADAF and NDAF. In
analytical calculations we take the adiabatic index $\gamma$ of the
inner disk to be 5/3 if the disk is gas pressure-dominated or
$\gamma=4/3$ if the disk is radiation or degeneracy
pressure-dominated. In numerical calculations, however, it is
convenient to introduce an ``equivalent" adiabatic index $\gamma$
based on the original definition of $\gamma$ from the first law of
thermodynamics, $\gamma=1+P/u$, where $P$ is the pressure of the
disk at a given $r$, and $u$ is the internal energy density at the
same radius. Therefore $\gamma$ is a variable as a function of
radius. We obtain the self-similar structure,
\begin{equation}
\frac{\rho(r)}{\rho(r+dr)}=\left(\frac{r}{r+dr}\right)^{-1/(\gamma(r)-1)},
\,\,\,\,
\frac{P(r)}{P(r+dr)}=\left(\frac{r}{r+dr}\right)^{-\gamma(r)/(\gamma(r)-1)}.\label{e34}
\end{equation}
If $\gamma$ does not vary significantly in the inner disk, the
difference between the approximate solution where $\gamma$ is a
constant and the numerical solution where we introduce an
``equivalent" $\gamma$ is not obvious.

In addition, some region of the accretion disk is optically thick to
neutrino emission when $\dot{m}_{d}$ is sufficiently large. We use
the same expressions of neutrino optical depth and emission at the
beginning of \S3. We also require that the neutrino emission
luminosity per unit area is continuous when the optical depth
crosses $\tau=1$.

We first calculate the value of $\tilde{r}$, which is the radius of
the boundary layer between the inner and outer disks. Figure 1 shows
$\tilde{r}$ as a function of $\dot{m}_{d}$ for different values of
$m$. We can see that when the disk flow is an ADAF at a low
accretion rate, $\tilde{r}$ decreases monotonously as the accretion
rate increases until the value of $\tilde{r}$ reaches a minimum. At
this minimum, the outer disk flow is an NDAF and most of the
neutrinos generated from the outer disk can escape freely, and the
effect of neutrino opacity is not important. If the accretion rate
is higher and makes the effect of neutrino opacity significant, the
value of $\tilde{r}$ increases with increasing the accretion rate.
From Figure 1, we see that when the accretion rate is either low or
sufficiently high, $\tilde{r}$ is very large and even reaches the
value of the outer radius, which means that the inner disk totally
covers the outer disk. In this situation, since the outer disk is
covered, no part of the disk is similar to that of the accretion
disk around a black hole as we discussed in $\S$3.1, and thus we say
that the entire disk becomes a self-similar structure and that the
physical variables of the entire disk are adjusted to build an
energy balance between heating and neutrino cooling. For this
situation, we do not want to discuss in details any more. We focus
on the accretion rate which allows the two-steady parts of disks to
exist.

In Figure 2, we choose several special conditions to plot the
density, temperature and pressure of the whole disk as functions of
radius $r$. Particularly, we take $\rho$, $P$ and $T$ to be
continuous at the boundary between the inner and outer disks. If the
parameter $\dot{m}_{d}$ is larger, or the disk contains more
neutrons (i.e., $Y_{e}$ becomes smaller), then the density,
temperature and pressure of the disk are larger, and the change of
the density and pressure is more dramatic than that of the
temperature. The change of the temperature cannot be very large
because it greatly affects the neutrino cooling rate of the disk.

If the mass of the central neutron star ($m$) becomes larger or the
electron fraction $Y_{e}$ becomes smaller, then the value of
$\tilde{r}$ in the monotonously decreasing segment of
$\dot{m}_{d}-\tilde{r}$ becomes smaller, and the value of
$\tilde{r}$ in the monotonously increasing segment of
$\dot{m}_{d}-\tilde{r}$ is larger. And the minimum value of
$\tilde{r}$ is almost independent of $Y_{e}$ and $m$. All of these
conclusions are consistent with the analytical solutions, except for
the case of the advection-dominated outer disk with the radiation
pressure to be dominated.  In $\S$3.2.2, we found that the size of
the inner disk increases with increasing $m$ for our analytical
solutions. However, by calculating the ``equivalent" adiabatic index
$\gamma$, we find that $\gamma$ decreases slightly with increasing
$m$. This makes the value of $\tilde{r}_{6}$ decrease, which is also
consistent with the analytical results (see Table 3). Figure 3 shows
the ``equivalent" adiabatic index $\gamma$ as a function of radius
of the entire disk for several different sets of parameters. In the
case where the accretion rate is low, the radiation pressure is
important. As the accretion rate increases, the gas pressure becomes
more dominant and the value of $\gamma$ is larger. On the other
hand, the ratio of the degeneracy pressure to the total pressure is
larger in the case of a higher accretion rate and $Y_{e}\sim 1$.
However, the gas pressure is always dominant for the accretion rate
chosen here. We see from Figure 3 that the change of $\gamma$ in the
inner disk is insignificant.

Figure 4 shows the ratio of the radial velocity $v_{r}$ and local
speed of sound $c_{s}$ as a function of $r$. The ratio is always
much smaller than unity, which means that the accretion flow is
always subsonic and no stalled shock exists in the disk. In many
cases the peak of this ratio is just at the boundary between the
inner and outer disks. The reason can be found in $\S$3.2.1, where
we gave the analytical expression of the ratio.
$f=1-(r_{*}/r)^{1/2}$ is the major factor that affects the value of
$v_{r}/c_{s}$ of the outer disk. However, in the inner disk,
$v_{r}/c_{s}$ always decreases since the isothermal sound speed can
be greater at smaller radii (i.e., $c_{s}\propto r^{-1/2}$), and the
radial velocity of the accreting gas, which satisfies the
self-similar solution (\ref{en02}), cannot change dramatically for
the disk matter to strike the neutron star surface.

Figure 5 shows the total neutrino emission luminosity of the entire
disk around a neutron star as a function of accretion rate for
parameters $M$ and $Y_{e}$ (where we do not consider neutrino
emission from the stellar surface discussed in \S 3.3), and we
compare it with the neutrino luminosity from a black-hole disk.
Also, we calculate the total neutrino luminosity from the inner and
outer disks. Here we roughly take the mass of the black hole to be
the same as that of the neutron star, and the innermost stable
circular orbit of the disk has a radius which is equal to the radius
of the neutron star. We approximately use the Newtonian dynamics for
simplicity. In Fig. 5, we find that the difference in neutrino
luminosity between the neutron-star and black-hole cases is a strong
function of the accretion rate. When the accretion rate is low
($\dot{m}_{d}\leq 10$), the total neutrino luminosity of the
black-hole disk $L_{\nu,\rm BH}$ is much smaller than that of the
neutron-star disk $L_{\nu,\rm NS}$, but $L_{\nu,\rm BH}$ and
$L_{\nu,\rm NS}$ are similar for a moderate accretion rate
($\dot{m}_{d}$ from 10 to 100). Actually, this result is consistent
with the general scenario introduced in \S2.1 and the basic result
shown in Fig. 1: for a low accretion rate, the black-hole disk is
mainly advection-dominated with most of the viscous
dissipation-driven energy to be advected into the event horizon of
the black hole, and we have $L_{\nu,\rm BH}\ll GM\dot{M}/(4r)$. On
the other hand, a large size of the inner disk of the neutron-star
disk for a low accretion rate makes the neutrino emission efficiency
be much higher than its black-hole counterpart. However, for a
moderate accretion rate, the black-hole disk is similar to the
neutron-star disk, which owns a quite small inner disk, and we have
$L_{\nu,\rm BH}\sim L_{\nu,\rm NS}$. Moreover, neutrino opacity
leads the value of $L_{\nu,\rm BH}$ to be less again compared with
$L_{\nu,\rm NS}$ for a high accretion rate, as this opacity
decreases the neutrino emission efficiency in the black-hole disk
but increases the size of the neutron-star disk again to balance the
heat energy release.

\section{An Elaborate Model of the Disk}
In the last section we first studied the disk structure
analytically. To do this, we used several approximations. First of
all, we took the pressure as a summation of several extreme
contributions such as the gas pressure of nucleons and electrons,
and the radiation pressure of a plasma of photons and $e^{+}e^{-}$
pairs. However, electrons may actually be degenerate or partially
degenerate, and the neutrino pressure should also be added to the
total pressure. Following Kohri et al. (2005), a lot of works about
hyperaccretion disks used the Fermi-Dirac distribution to calculate
the pressure of electrons and even the pressure of nucleons. Second,
neutrino cooling we used in the last section is simplified,
following Popham et al. (1999) and Narayan et al. (2001) and
neglecting the effect of electron degeneracy and the effect of
different types of neutrinos and their different optical depth. In
fact, these effects may be significant in some cases. Third, we took
the electron-nucleon-radio $Y_{e}$ as a constant parameter in our
analytic model in $\S$3. Realistically, $Y_{e}$ should be calculated
based on $\beta$-equilibrium and neutronization in hyperaccretion
disks. In this section, we still use the assumption of outer and
inner disks discussed in $\S$2, but consider a state-of-the-art
model with lots of elaborate (more physical) considerations on the
thermodynamics and microphysics in the disk, which was recently
developed in studying the neutrino-cooled disk of a black hole. In
addition, we compare results from this elaborate model with those of
the simple model discussed in $\S$3.

\subsection{Thermodynamics and microphysics}
The total pressure in the disk can be written as: $P=P_{\rm
nuc}+P_{\rm rad}+P_{e}+P_{\nu}$. We still consider all the nucleons
to be free ($X_{\rm nuc}\approx 1$) as mentioned in \S 2.2, and
ignore the photodisintegration process. Also, we replace the term of
radiation pressure $11aT^{4}/12$ in the simple model by $aT^{4}/3$
in this section, because the pressure of $e^{+}e^{-}$ pairs can be
calculated in the electron pressure $P_{e}$ with the Fermi-Dirac
distribution:
\begin{equation}
P_{e^{\pm}}=\frac{1}{3}\frac{m_{e}^{4}c^{5}}{\pi^{2}\hbar^{3}}
\int^{\infty}_{0}\frac{x^{4}}{\sqrt{x^{2}+1}}\frac{dx}{e^{(m_{e}c^{2}\sqrt{x^{2}+1}\mp\mu_{e})/k_{B}T}+1},\label{z02}
\end{equation}
where $x=p/m_ec$ is the dimensionless momentum of an electron and
$\mu_{e}$ is the chemical potential of the electron gas. $P_{e}$ is
the summation of $P_{e^{-}}$ and $P_{e^{+}}$. In addition, we take
the neutrino pressure to be
\begin{equation}
P_{\nu}=u_{\nu}/3,\label{z03}
\end{equation}
where $u_{\nu}$ is the energy density of neutrinos.

The ``equivalent" adiabatic index can be expressed by
\begin{equation}
\gamma=1+(P_{\rm nuc}+P_{\rm rad}+P_{e}+P_{\nu})/(u_{\rm nuc}+u_{\rm
rad}+u_{e}+u_{\nu}). \label{z061}
\end{equation}
with the inner energy density to be
\begin{equation}
u_{\rm gas}=\frac{3}{2}P_{\rm gas},\label{z04}
\end{equation}
\begin{equation}
u_{\rm rad}=3P_{\rm rad},\label{z05}
\end{equation}
\begin{equation}
u_{e^{\pm}}=\frac{m_{e}^{4}c^{5}}{\pi^{2}\hbar^{3}}
\int^{\infty}_{0}\frac{x^{2}\sqrt{x^{2}+1}}{e^{(m_{e}c^{2}\sqrt{x^{2}+1}\mp\mu_{e})/k_{B}T}+1}dx.\label{z06}
\end{equation}
We then use equation (\ref{e34}) to obtain the self-similar inner
disk.

In addition, we add the equation of charge neutrality among protons,
electrons and positrons to estabilish the relation between $\rho$,
$Y_{e}$ and $\mu_{e}$.
\begin{equation}
n_{p}=\frac{\rho Y_{e}}{m_{B}}=n_{e^{-}}-n_{e^{+}},\label{z062}
\end{equation}
where we use the Fermi-Dirac form to calculate $n_{e^{-}}$ and
$n_{e^{+}}$.

Moreover, in the elaborate model, we adopt the improved formula of
the neutrino cooling rate $Q_{\nu}^{-}$, the inner energy density of
neutrinos $u_{\nu}$ , as well as the absorption and scattering
optical depth for three types neutrinos
$\tau_{a,\nu_{i}(e,\mu,\tau)}$ and $\tau_{s,\nu_{i}(e,\mu,\tau)}$
following a series of previous work (e.g., Popham \& Narayan 1995,
Di Matteo et al. 2002, Kohri et al. 2005, Gu et al. 2006, Janiuk et
al. 2007 and Liu et al. 2007). The three types of neutrino cooling
rate per unit volume are
\begin{equation}
\dot{q}_{\nu_{e}}=\dot{q}_{\rm eN}+\dot{q}_{e^{-}e^{+}\rightarrow
\nu_{e}\bar{\nu}_{e}}+\dot{q}_{\rm brem}+\dot{q}_{\rm
plasmon},\label{z09}
\end{equation}
\begin{equation}
\dot{q}_{\nu_{\mu}}=\dot{q}_{\nu_{\tau}}=\dot{q}_{e^{-}e^{+}\rightarrow
\nu_{\tau}\bar{\nu}_{\tau}}+\dot{q}_{\rm brem}, \label{z091}
\end{equation}
where the meanings of four terms $\dot{q}_{\rm eN}$,
$\dot{q}_{e^{-}e^{+}\rightarrow \nu_{i}\bar{\nu}_{i}}$,
$\dot{q}_{\rm brem}$, and $\dot{q}_{\rm plasmon}$ have been shown at
the beginning of $\S$3. Here $\dot{q}_{\rm eN}$ and $\dot{q}_{\rm
plasmon}$ are only related to $\dot{q}_{\nu_{e}}$. Moreover,
$\dot{q}_{\rm eN}$ is a summation of three terms,
\begin{equation}
\dot{q}_{\rm eN}=\dot{q}_{p+e^{-}\rightarrow
n+\nu_{e}}+\dot{q}_{n+e^{+}\rightarrow
p+\bar{\nu}_{e}}+\dot{q}_{n\rightarrow
p+e^{-}+\bar{\nu}_{e}}.\label{z092}
\end{equation}
The formulae of three terms in equation (\ref{z092}) are the same as
Kohri et al. (2005), Janiuk et al. (2007) and Liu et al. (2007), who
considered the effect of electron degeneracy. In addition, we use
the same formulae of $\dot{q}_{e^{-}e^{+}\rightarrow
\nu_{i}\bar{\nu}_{i}}$, $\dot{q}_{\rm brem}$, and $\dot{q}_{\rm
plasmon}$ as the early works such as Kohri et al. (2002, 2005) and
Liu et al. (2007).

Finally, different from the simple model in $\S$3 in which we took
the electron fraction $Y_{e}$ as a free parameter, in this section
we calculate $Y_{e}$ by considering the $\beta$-equilibrium in the
disk among electrons and nucleons following Lee et al. (2005) and
Liu et al. (2007)
\begin{equation}
\textrm{ln}\left(\frac{n_{n}}{n_{p}}\right)
=f(\tau_{\nu})\frac{2\mu_{e}-Q}{k_{B}T}+[1-f(\tau_{\nu})]\frac{\mu_{e}-Q}{k_{B}T},\label{z10}
\end{equation}
with the weight factor $f(\tau_{\nu})$=exp$(-\tau_{\nu_{e}})$ and
$Q=(m_{n}-m_{p})c^{2}$. Equation (\ref{z10}) is a combined form to
allow the transition from the neutrino-transparent limit case to the
neutrino-opaque limit case of the $\beta$-equilibrium.

\subsection{Numerical results in the elaborate model}
Using equations (\ref{e1}), (\ref{e2}), (\ref{e3}) and (\ref{e31})
in $\S$2.2 and the improved treatment in $\S$4.1, we can solve the
structure of the outer disk. Then using equations (\ref{en02}) and
(\ref{e4}) in $\S$2.3, we determine the size and the structure of
the inner disk, and calculate the neutrino luminosity of the entire
disk.

Figure 6a shows the size of the inner disk in the elaborate model.
We still choose the mass of the central neutron star to be
$M=1.4M_{\odot}$ and $M=2.0M_{\odot}$. From Figure 6a we can see
that the size of the inner disk $\tilde{r}$ decreases with
increasing the accretion rate, and reaches a minimum at $\dot{M}
\sim 0.1 M_{\odot}s^{-1}$. Then the value of $\tilde{r}$ increases
with increasing the accretion rate. This result is well consistent
with what we have found in the simple model in \S 3. The physical
reason for this result has been discussed in \S 3.3.2 and \S 3.4.
Figure 6b shows the solution of the inner disk size both in the
simple model and the elaborate model. We fix the central neutron
star $M=1.4M_{\odot}$. In the simple model of $\S$3, we take $Y_{e}$
as a free parameter and plot two $\dot{m}_{d}-\tilde{r}_{6}$ lines
with $Y_{e}=1$ and $Y_{e}=1/9$, while in the elaborate model we only
plot one line since $Y_{e}$ can be directly determined through
$\beta$-equilibrium in the disk. From Figure 6b we conclude that the
solutions of the two models are basically consistent with each
other.

In Figure 7, we shows the ``equivalent" adiabatic index $\gamma$ and
the electron fraction $Y_{e}$ in the entire disk for three values of
the accretion rate $\dot{M}=0.01M_{\odot}s^{-1}$,
$0.1M_{\odot}s^{-1}$ and $1.0M_{\odot}s^{-1}$, and for two values of
the mass of the central neutron star $M=1.4M_{\odot}$ and
$2.0M_{\odot}$. The ``equivalent" adiabatic index $\gamma$ increases
with increasing the accretion rate in most region of the disks,
since gas will take over electrons and radiation to be the dominant
pressure when the accretion rate is high enough. In addition,
$\gamma$ decreases as the radius decreases in the inner disk. These
are consistent with the results of the simple model (see Fig. 3).
From Figure 7b, we can see that $Y_{e}\sim 1$ when the accretion
rate is low, and $Y_{e}\ll 1$ when the accretion rate becomes
sufficiently high. This result is consistent with Kohri et al.
(2005, their Fig. 6b). Chen \& Beloborodov (2007) and Liu et al.
(2007) showed the electron fraction $Y_{e}\leq 0.5$ in the disk,
since they supposed that initial neutrons and protons come from
photodisintegration of $\alpha$-particles at some large radius far
from a central black hole. However, since the hyperaccretion disk
around a neutron star we discuss has a size smaller than that of a
black-hole disk, we consider the mass fraction of free nucleons
$X_{\rm nuc}=1$ in the entire disk. Therefore, the fraction of
protons can be slightly higher and it is possible that the protons
are richer than neutrons in the disk or $Y_{e}\geq 0.5$ if the
accretion rate is low enough.

Figure 8 and Figure 9a show the density, temperature, pressure and
the neutrino luminosity per unit area in the entire disk with three
different accretion rates. We fix $M=1.4M_{\odot}$, and also plot
two other curves of solutions in the simple model with $Y_{e}=1$ and
$Y_{e}=1/9$. The density $\rho$ and pressure $P$ in the elaborate
model are smaller than those in the simple model when the accretion
rate is low ($\dot{m}_{d}=1.0$), but are similar to the solution of
the simple model with $Y_{e}=1/9$ in the high accretion rate
($\dot{m}_{d}=100$). In addition, $\rho$ and $P$ in the elaborate
model change from one solution ($Y_{e}=1$) to another solution
($Y_{e}=1/9$) in the simple model for an intermediate accretion rate
(e.g., $\dot{m}_{d}\sim 10$), since $Y_{e}\sim 0.5$ in the outer
edge of the disk and $Y_{e}\ll 1$ in the inner disk. The
distribution of the neutrino cooling rate $Q^{-}_{\nu}$ (luminosity
per unit area) in the elaborate model is almost the same as that in
the simple model with $Y_{e}=1$ and low accretion rate or
$Y_{e}=1/9$ and a high accretion rate. However, the value of
$Q^{-}_{\nu}$ is still different in these two models for the region
that is optically thick to neutrino emission in the disk.

We also plot the the total neutrino emission luminosity of the
entire disk, the outer and inner disks as functions of accretion
rate with the central neutron star mass of $1.4M_{\odot}$, and
compare the total neutrino luminosity with that of the black-hole
disk (Figure 9b). The results are similar to what we have found in
the simple model (Figure 5).

\section{Conclusions and Discussions}
In this paper we have studied the structure, energy advection and
conversion, and neutrino emission of a hyperaccretion disk around a
neutron star. We considered a quasi-steady disk model without any
outflow. Similar to the disk around a black hole, the neutron star
disk with a huge mass accretion rate is extremely hot and dense,
opaque to photons, and thus is only cooled via neutrino emission, or
even optically thick to neutrino emission in some region of the disk
if the accretion rate is sufficiently high. However, a significant
difference between black hole and neutron star disks is that the
heat energy of the disk can be advected into the event horizon if
the central object is a black hole, but if the central object is a
neutron star, the heat energy should be eventually released from a
region of the disk near the stellar surface. As a result, the
neutrino luminosity of the neutron star disk should be much larger
than that in black hole accretion. We approximately took the disk as
a Keplerian disk. According to the Virial theorem, one half of the
gravitational energy in such a disk is used to heat the disk and the
other half to increase the rotational kinetic energy of the disk. We
assumed that most of the heat energy generated from the disk is
still cooled from the disk via neutrino emission and the rotational
energy is used to spin up the neutron star or is released on the
stellar surface via neutrino emission.

In a certain range of hypercritical accretion rates, depending on
the mechanisms of energy heating and cooling in the disk, we
considered a two-region, steady-state disk model. The outer disk is
similar to the outer region of a black hole disk. We used the
standard viscosity assumption, Newtonian dynamics and vertically
integrated method to study the structure of the outer disk. Since
the radial velocity of the disk flow is always subsonic, no stalled
shock exists in the disk and thus we considered that physical
variables in the disk change continuously when crossing the boundary
layer between the inner and outer disks. The inner disk, which
expands until a heating-cooling balance is built, could satisfy a
self-similar structure as shown by equation (\ref{en02}).

In this paper we first studied the disk structure analytically. To
do this, we adopted a simple disk model based on the analytical
method. We took the pressure as a summation of several extreme
contributions and simple formulae of neutrino cooling. And we took
the electron fraction $Y_{e}$ as a parameter in the simple model. We
used an analytical method to find the dominant-pressure distribution
(Table 2) and the radial distributions of the density, temperature
and pressure (solutions \ref{s11}, \ref{s12}, \ref{e302}, \ref{s2},
\ref{e3031}) in the outer disk. Then we used the equation of energy
balance between heating and neutrino cooling to calculate the size
of the inner disk in four different cases: whether the
advection-dominated outer disk is radiation or gas
pressure-dominated, and whether the neutrino-cooled outer disk is
optically thin or thick to neutrino emission (Table 3 to 6).
Subsequently, we numerically calculated the size of the inner disk,
the structure, and energy conversion and emission of the entire disk
in the simple model (Fig. 2 to Fig. 5) and compared the numerical
results with the analytical results. The numerical results are
consistent with the analytical ones from the simple model.

When the accretion rate is sufficiently low, most of the disk is
advection-dominated, the energy is advected inward to heat the inner
disk, and eventually released via neutrino emission in the inner
disk. In this case, the inner disk is very large, and quite
different from a black hole disk, which advects most of the energy
inward into the event horizon. If the accretion rate is higher, then
physical variables such as the density, temperature and pressure
become larger, the disk flow becomes NDAF, the advected energy
becomes smaller, and heating of the inner disk becomes less
significant. As a result, the size of the inner disk is much
smaller, and the difference between the entire disk and the black
hole disk becomes less significant. Furthermore, if the accretion
rate is large enough to make neutrino emission optically thick, then
the effect of neutrino opacity becomes important so that the
efficiency of neutrino emission from most of the disk decreases and
the size of the inner disk again increases until the entire disk
becomes self-similar. Besides, a different mass of the central star
or a different electron-nucleon ratio also makes physical variables
and properties of the disk different. However, the accretion rate
plays a more significant role in the disk structure and energy
conversion, as it varies much wider than the other parameters.

The simple model is based on the early works such as Popham et al.
(1999) and Narayan et al. (2001). We found that the simple model in
fact gives us a clear physical picture of the hyperaccretion disk
around a neutron star, even if we used some simplified formulae in
thermodynamics and microphysics in the disk. In $\S$4 we considered
an elaborate model, in which we calculated the pressure of electrons
and positrons by using the Fermi-Dirac distribution and replaced the
factor $11/12$ by $1/3$ in the radiation-pressure equation. We
adopted more advanced expressions of the neutrino cooling rates,
including the effect of all three types of neutrinos and the
electron degeneracy. Moreover, we considered $\beta$-equilibrium in
the disk to calculate the electron fraction $Y_{e}$. Then, in the
elaborate model, we also calculated the the size of the inner disk
(Fig. 6), the radial distributions of the ``equivalent" adiabatic
index $\gamma$, the electron fraction (Fig. 7), the density,
temperature and pressure in the disk (Fig. 8), the neutrino cooling
rate distribution of the disk (Fig. 9a), the neutrino emission
luminosity from the inner and outer disks, and the total neutrino
luminosity of a neutron-star disk compared with that of a black-hole
disk (Fig. 9b).

The electron fraction $Y_{e}$ was also determined in the elaborate
model. We found that $Y_{e}$ drops with increasing the accretion
rate in the outer disk. $Y_{e}$ can be greater than 0.5 at a large
radius if the accretion rate is sufficiently low, and $Y_{e}\ll 1$
in the disk when the accretion rate is high enough. If we put these
results of $Y_{e}$ in the elaborate model into the simple model
correctly (i.e. $Y_{e}\sim 1$ for low accretion rate and $Y_{e}\ll
1$ for high accretion rate), we find that they are basically
consistent with each other (see Fig. 6, Fig. 7, Fig. 8, Fig. 9), and
also consistent with most of the early works (see the discussion in
the end of \S3.1).

A main difference in the structure between the simple and elaborate
models is caused by different expressions of pressure adopted in the
two models. In order to see it clearly, we introduce the third model
here. We still keep $P_{\rm rad}=11aT^{4}/12$ as the simple model
but just change the relativistic degeneracy pressure term of
electrons (the second term in equation \ref{e32}) to the Fermi-Dirac
distribution (formula \ref{z02} for $P_{e^{-}}$), and use all the
other formulae in $\S$4. In other words, the third model is
introduced by only changing one pressure term in the elaborate
model. We can find that the results from the third model are even
more consistent with those of the simple model than the elaborate
model in \S 4. Take Figure 10 as an example. We compare the solution
of the inner disk of the third model with that of the simple model.
From Fig. 10 we can see that if the accretion rate is low and
$Y_{e}\sim 1$, the thick solid line is much closer to the thin
dashed line, which results from the simple model with $Y_{e}=1$; and
if the accretion rate is high enough and $Y_{e}\ll 1$, the solid
line is much closer to the thick dotted line, which is the result
from the simple model with $Y_{e}=1/9$. Compared with Fig. 8a, this
result is even more consistent with the simple model. The values of
density $\rho$ and pressure $P$ in Fig. 10, for a low accretion
rate, are smaller than those of the simple model, which is also due
to different expressions of the radiation pressure used in these two
models in \S 3 and \S 4. Therefore, we conclude that the main
difference of the results between the simple model in \S 3 and the
elaborate model in \S 4 come from different expressions of the
pressure in disks. However, we believe that the pressure formulae
given in the elaborate model are more realistic since $11aT^{4}/12$
is only an approximated formula for the pressure of a plasma of
photons and $e^{+}e^{-}$ pairs. On the other hand, as what has been
pointed out by Lee et al. (2005) and Liu et al. (2007), formulae
$P_{\rm rad}=aT^{4}/3$ and (\ref{z02}) in \S 4.1.1 are better and
can automatically take relativistic $e^{+}e^{-}$ pairs into account
in the expression of $P_{e}$.

The different expression of the neutrino cooling rate $Q^-_{\nu}$
makes the neutrino luminosity distribution different in the region
where is optically thick to neutrino emission. A more advanced
expression of neutrino cooling rate $Q_{\nu}^{-}$ gives better
results of the neutrino luminosity per unit area than that given by
the rough expression $\frac{7}{8}\sigma_{B}T^{4}/\tau$ in \S3.

In this paper we studied the disk without any outflow, which may
exist if the disk flow is an ADAF. However, it is still unclear
whether an outflow or neutrino cooling plays a more important role
since the size of the disk is quite small. The other case that we
ignored is that, if the radius of the central neutron star is
smaller than that of the innermost stable circular orbit of the
accretion disk, the accreting gas eventually falls onto the neutron
star freely. In this case, a shock could form in the region between
the innermost stable circular orbit and neutron star surface
(Medvedev 2004). This effect can be studied if other effects such as
the equation of state of a differentially-rotating neutron star and
its mass-radius relation are together involved.

Neutrinos from a hyperaccretion disk around a neutron star will be
possibly annihilated to electron/positron pairs, which could further
produce a jet. It would be expected that such a jet is more
energetic than that from the neutrino-cooled disk of a black hole
with same mass and accretion rate as those of the neutron star
(Zhang \& Dai 2008, in preparation). This could be helpful to draw
the conclusion that some GRBs originate from neutrino annihilation
rather than magnetic effects such as the Blandford-Znajek effect.

We considered a central neutron star with surface magnetic field
weaker than $B_{s,{\rm cr}}\sim 10^{15}-10^{16}$ G for typical
hyperaccretion rates in this paper. For magnetars (i.e., neutron
stars with ultra-strongly magnetic fields of $\sim B_{s,{\rm cr}}$),
however, the magnetic fields could play a significant role in the
global structure of hyperaccretion disks as well as underlying
microphysical processes, e.g., the quantum effect (Landau levels) on
the electron distribution and magnetic pressure in the disks could
become important. Thus, the effects of an ultra-strongly magnetic
field on hyperaccretion disks around neutron stars are an
interesting topic, which deserves a detailed study.

\section*{Acknowledgements}
We would like to thank the referees for their valuable comments that
have allowed us to improve this manuscript. We also thank P. F. Chen
for his helpful discussion in numerical solutions and Yefei Yuan and
Bing Zhang for their useful suggestions. This work is supported by
the National Natural Science Foundation of China (grants 10221001
and 10640420144) and the National Basic Research Program of China
(973 program) No. 2007CB815404.

\newpage
\begin{table}
\begin{tabular}{|l|l|r|r|r|r|}
\hline
notation & definition & \S/Eq. \\
\hline
 $m$ & mass of the central neutron star, $m=M/1.4M_{\odot}$ & \S3.1, (\ref{s11})\\
 $\dot{m}_{d}$ & mass accretion rate, $\dot{m}_d=\dot{M}/0.01M_{\odot}\,{\rm s}^{-1}$ & \S3.1, (\ref{s11})\\
 $Y_{e}$ & ratio of the electron to nucleon number density in the disk & \S2.2, (\ref{a02})\\
 $r_{*}$ & radius of the neutron star & \S2.3, (\ref{e231}) \\
 $r_{\rm out}$ & outer radius of the disk & \S2.3, (\ref{e231}) \\
 $\Omega_{*}$ & angular velocity of the stellar surface & \S3.3, (\ref{e331}) \\
 $\varepsilon$ & efficiency of energy release in the inner disk & \S2.3, (\ref{e4}) \\
 $\tilde{r}$ & radius between the inner and outer disks & \S2.3, (\ref{en02}) \\
 $\bar{r}$ & radius between the neutrino optically-thick \& -thin regions & \S3.2, (\ref{e10}) \\
 $k=\bar{r}/\tilde{r}$ & parameter to measure the neutrino optically-thick region & --- \\
 $f=1-\sqrt{\frac{r_{*}}{r}}$ & useful factor as a function of $r$& \S2.2, (\ref{e21}) \\
 $\tilde{f}=1-\sqrt{\frac{r_{*}}{\tilde{r}}}$ & value of $f$ at radius $\tilde{r}$ & \S3.2, (\ref{e11}) \\
 $\bar{f}_{\nu}$ & average efficiency of neutrino cooling in the outer disk & \S2.3, (\ref{e231}) \\
\hline
\end{tabular}
\caption{Notation and definition of some quantities in this paper.}
\end{table}
\begin{table}
\begin{center}
\begin{tabular}{rccccrc}
\hline \hline
dominant pressure & accretion flow & range of $\dot{m}_{d}$ & $Y_{e}\sim 1$ & $Y_{e}\ll 1$ \\
\hline
$P_{\rm rad}$....................... & ADAF & (a) \& (b) & 0.453 & 7.25\\
$P_{\rm gas}$....................... & ADAF & (c) \& (d) & 1.73 & ---\\
$P_{\rm gas}$....................... & NDAF & (e) \& (f) & 64.5 & $\infty$\\
$P_{\rm deg}$....................... & NDAF & (g) & $\infty$ & ---\\
\hline \hline
\end{tabular}
\caption{Range of the accretion rate in different regions. (a)
$\dot{m}_{d}<0.504m^{1/2}\alpha_{-1}^{5/3}r_{6}^{5/2}f^{1/6}$; (b)
$\dot{m}_{d}<644(1+Y_{e})^{-4}m^{7/2}\alpha_{-1}r_{6}^{-3/2}f^{7/2}$;
(c)
$\dot{m}_{d}>644(1+Y_{e})^{-4}m^{7/2}\alpha_{-1}r_{6}^{-3/2}f^{7/2}$;
(d)
$\dot{m}_{d}<4.81\times10^{-3}(1+Y_{e})^{13/5}m^{-29/20}\alpha_{-1}^{21/10}r_{6}^{47/20}f^{-2}$;
(e)
$\dot{m}_{d}>4.81\times10^{-3}(1+Y_{e})^{13/5}m^{-29/20}\alpha_{-1}^{21/10}r_{6}^{47/20}f^{-2}$;
(f)
$\dot{m}_{d}<0.148Y_{e}^{-4}(1+Y_{e})^{27/5}m^{-11/10}\alpha_{-1}^{19/10}r_{6}^{33/20}f^{-1}$;
(g)
$\dot{m}_{d}>0.148Y_{e}^{-4}(1+Y_{e})^{27/5}m^{-11/10}\alpha_{-1}^{19/10}r_{6}^{33/20}f^{-1}$.}
\end{center}
\end{table}
\begin{table}
\begin{center}
\begin{tabular}{rccccrc}
\hline \hline
$\dot{m}_{d}$ & 0.01 & 0.05 & 0.1 & 0.5 \\
\hline
Case 1 & 6.38 & 3.53 & 2.76 & 1.65 \\
Case 2 & 13.87 & 5.66 & 3.92 & 1.84 \\
Case 3 & 6.83 & 3.76 & 2.94 & 1.74 \\
\hline \hline
\end{tabular}
\caption{Equation (\ref{e6}) gives $\tilde{r}_{6}$ in several cases.
Case 1: $m=1, \gamma=4/3$; case 2: $m=1, \gamma=1.4$; case 3:
$m=2.0/1.4, \gamma=4/3$.}
\end{center}
\end{table}
\begin{table}
\begin{center}
\begin{tabular}{rccccrc}
\hline \hline
$\dot{m}_{d}$ & 0.5 & 1 & 3 & 5 & 10 \\
\hline
Case 1 & 2.17 & 1.91 & 1.62 & 1.52 & 1.42\\
Case 2 & 2.05 & 1.83 & 1.59 & 1.50 & 1.40\\
Case 3 & 1.68 & 1.54 & 1.38 & 1.33 & 1.27\\
Case 4 & 1.97 & 1.76 & 1.52 & 1.44 & 1.36\\
\hline \hline
\end{tabular}
\caption{Equation (\ref{e321}) gives $\tilde{r}_{6}$ in several
cases. Case 1: $Y_{e}=1, \gamma=5/3, m=1$; case 2: $Y_{e}=1,
\gamma=3/2, m=1$; case 3: $Y_{e}=1/9, \gamma=5/3, m=1$; case 4:
$Y_{e}=1, \gamma=5/3, m=2.0/1.4$.}
\end{center}
\end{table}

\begin{table}
\begin{center}
\begin{tabular}{rccccrc}
\hline \hline
$\bar{f}_{\nu}$$\setminus$$\gamma$ & 5/3 & 3/2 & 4/3 \\
\hline
0.9 & 1.31 & 1.28 & 1.26 \\
0.7 & 1.46 & 1.43 & 1.39 \\
0.5 & 1.56 & 1.53 & 1.47 \\
\hline \hline
\end{tabular}
\caption{Equation (\ref{e7}) gives $\tilde{r}_{6}$. We take
$\bar{f}_{\nu}$ and $\gamma$ as parameters to solve this equation.}
\end{center}
\end{table}

\begin{table}
\begin{center}
\begin{tabular}{rccccrc}

\hline \hline
$\dot{m}_{d}$ & 80 & 100 & 120 & 150 \\
\hline
Case 1 & 4.28 & 5.13 & 5.94 & 7.10 \\
Case 2 & 6.26 & 7.48 & 8.65 & 10.33 \\
Case 3 & 5.09 & 6.09 & 7.05 & 8.43 \\
Case 4 & 4.10 & 4.99 & 5.86 & 7.11 \\
Case 5 & 4.87 & 5.90 & 6.89 & 8.32 \\
\hline \hline
\end{tabular}
\caption{Equation (\ref{e8}) gives $\tilde{r}_{6}$ in several
different cases. Case 1: $Y_{e}=1, m=1, k=1, \gamma=5/3$; case 2:
$Y_{e}=1/ 9, m=1, k=1, \gamma=5/3$; case 3: $Y_{e}=1, m=2.0/1.4,
k=1, \gamma=5/3$, Case 4: $Y_{e}=1, m=1, k=0.7, \gamma=5/3$; case 5:
$Y_{e}=1, m=1, k=1, \gamma=3/2$. }
\end{center}
\end{table}

\newpage
\begin{figure}
\resizebox{\hsize}{!} {\includegraphics{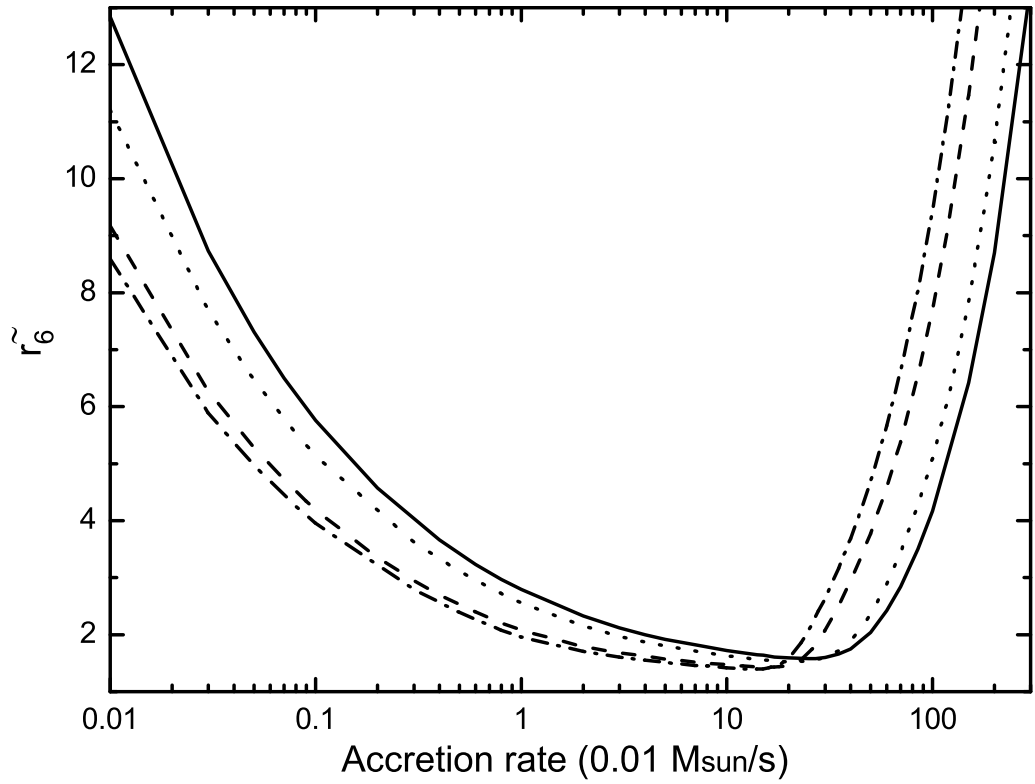}} \caption{The
radius $\tilde{r}_{6}$ of the boundary layer between the inner and
outer disks as a function of accretion rate $\dot{m}_{d}$ in the
simple model for several sets of parameters: (1) $M=1.4M_{\odot}$
and $Y_{e}=1.0$ ({\em solid line}), (2) $M=1.4M_{\odot}$ and
$Y_{e}=1/9$ ({\em dashed line}), (3) $M=2.0M_{\odot}$ and
$Y_{e}=1.0$ ({\em dotted line}), and (4) $M=2.0M_{\odot}$ and
$Y_{e}=1/9$ ({\em dash-dotted line}).}
\end{figure}

\newpage
\begin{figure}
\resizebox{\hsize}{!} {\includegraphics{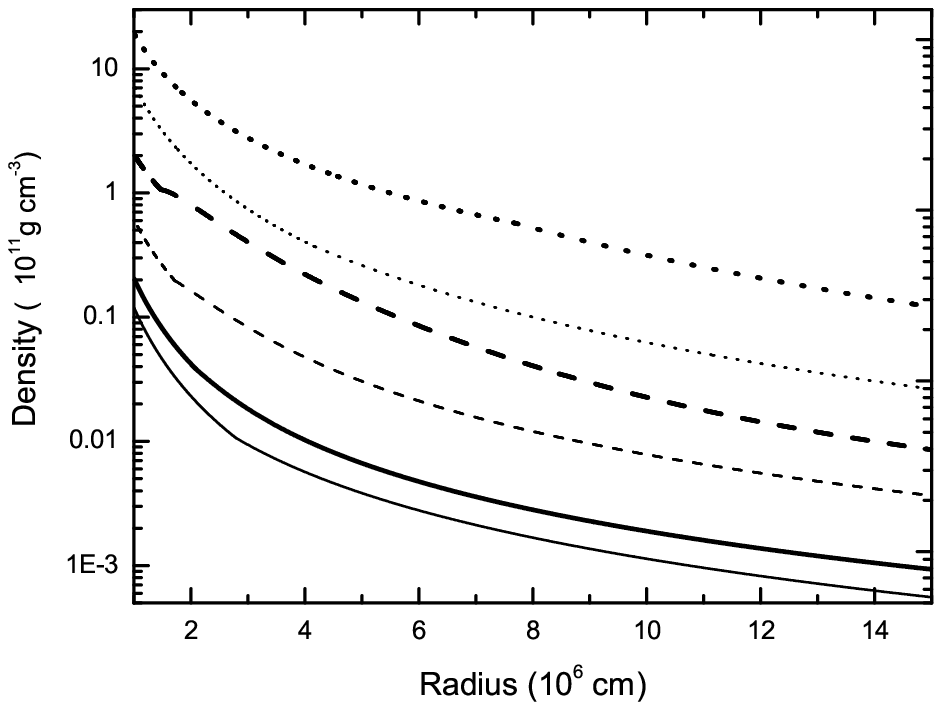}
\includegraphics{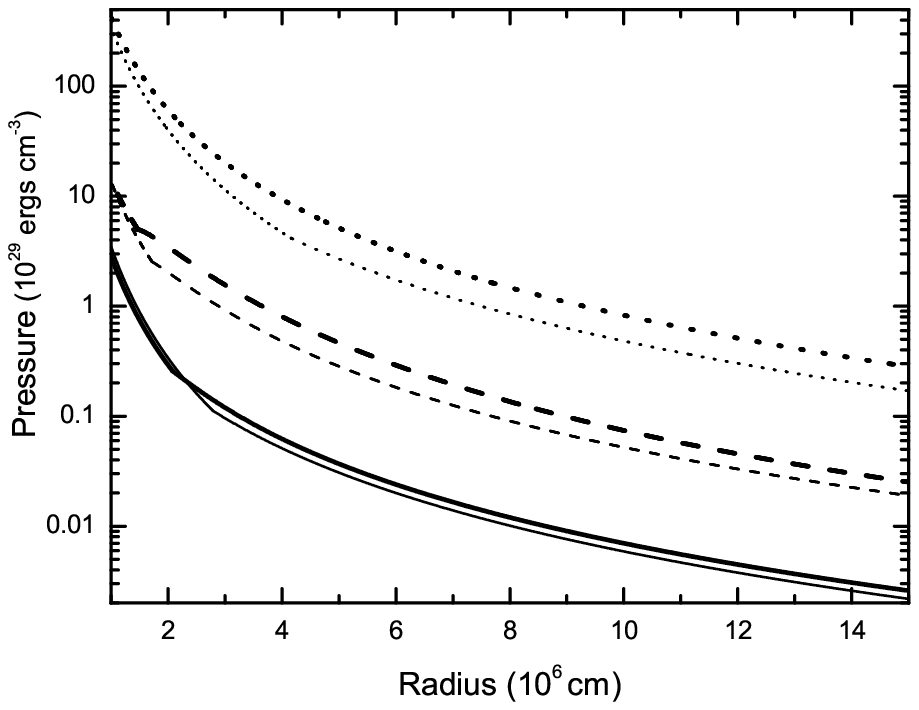}}
\\\resizebox{\hsize}{!} {\includegraphics{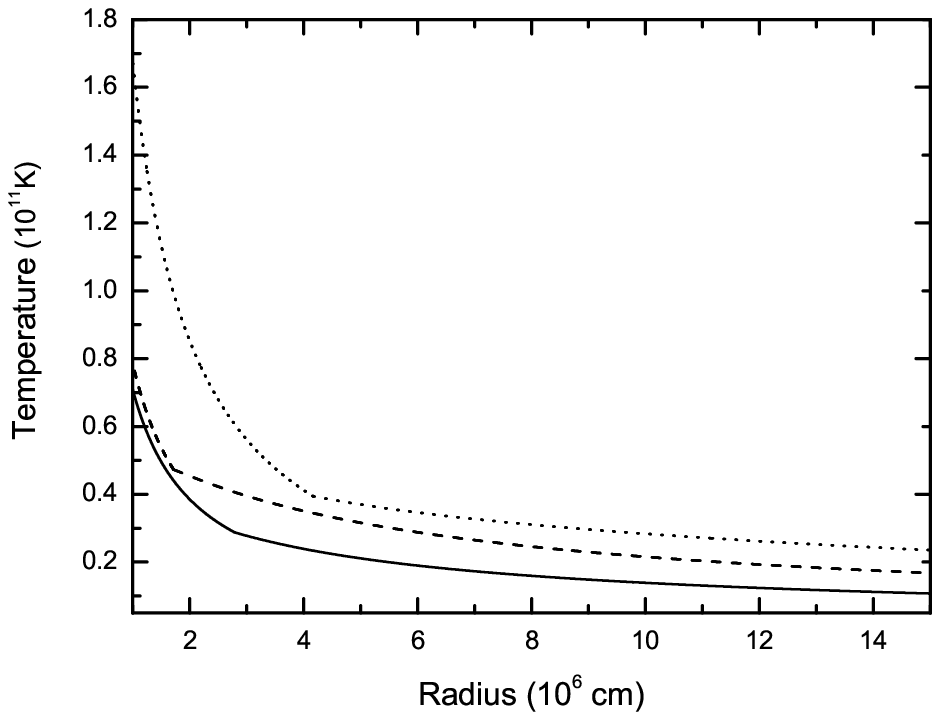}
\includegraphics{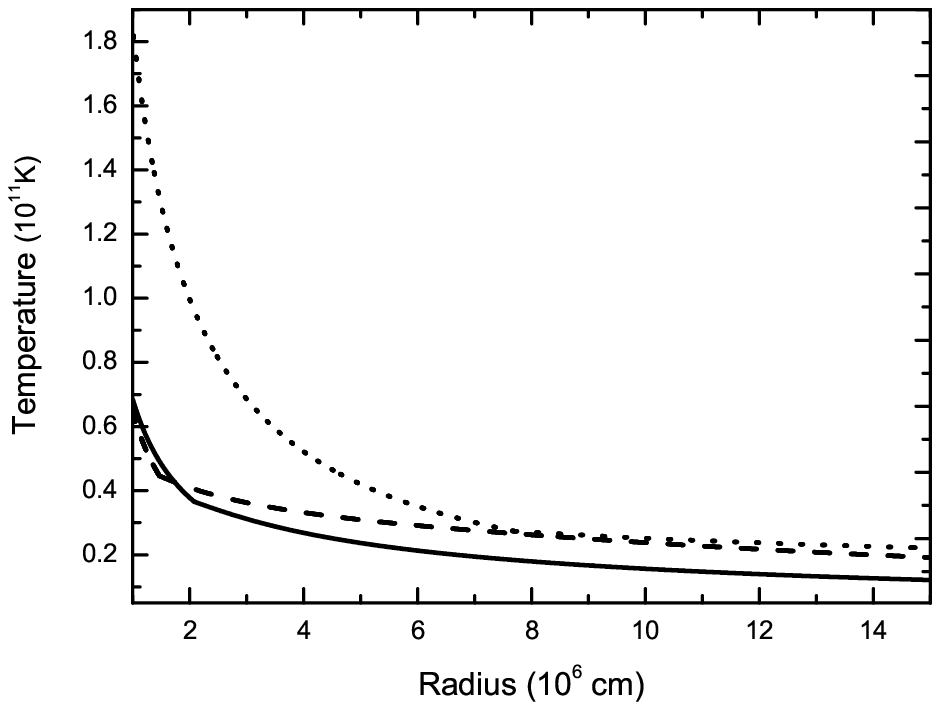}}
\caption{The density (in units of $10^{11}$g cm$^{-3}$), pressure
(in units of $10^{11}$ ergs cm$^{-3}$) and temperature (in units of
$10^{11}$K) of the disk as functions of radius $r$ (in units of
$10^{6}$ cm) in the simple model for several sets of parameters: (1)
$M=1.4M_{\odot}$, $Y_{e}=1.0$, and $\dot{M}=0.01M_{\odot}\,{\rm
s}^{-1}$ ({\em thin solid line}), (2) $M=1.4M_{\odot}$, $Y_{e}=1.0$,
and $\dot{M}=0.1M_{\odot}\,{\rm s}^{-1}$ ({\em thin dashed line}),
(3) $M=1.4M_{\odot}$, $Y_{e}=1.0$, and $\dot{M}=1.0M_{\odot}\,{\rm
s}^{-1}$ ({\em thin dotted line}), (4) $M=2.0M_{\odot}$,
$Y_{e}=1.0$, and $\dot{M}=0.01M_{\odot}\,{\rm s}^{-1}$ ({\em thick
solid line}), (5) $M=2.0M_{\odot}$, $Y_{e}=1.0$, and
$\dot{M}=0.1M_{\odot}\,{\rm s}^{-1}$ ({\em thick dashed line}), and
(6) $M=2.0M_{\odot}$, $Y_{e}=1.0$, and $\dot{M}=1.0M_{\odot}\,{\rm
s}^{-1}$ ({\em thick dotted line}). }
\end{figure}

\newpage
\begin{figure}
\resizebox{\hsize}{!} {\includegraphics{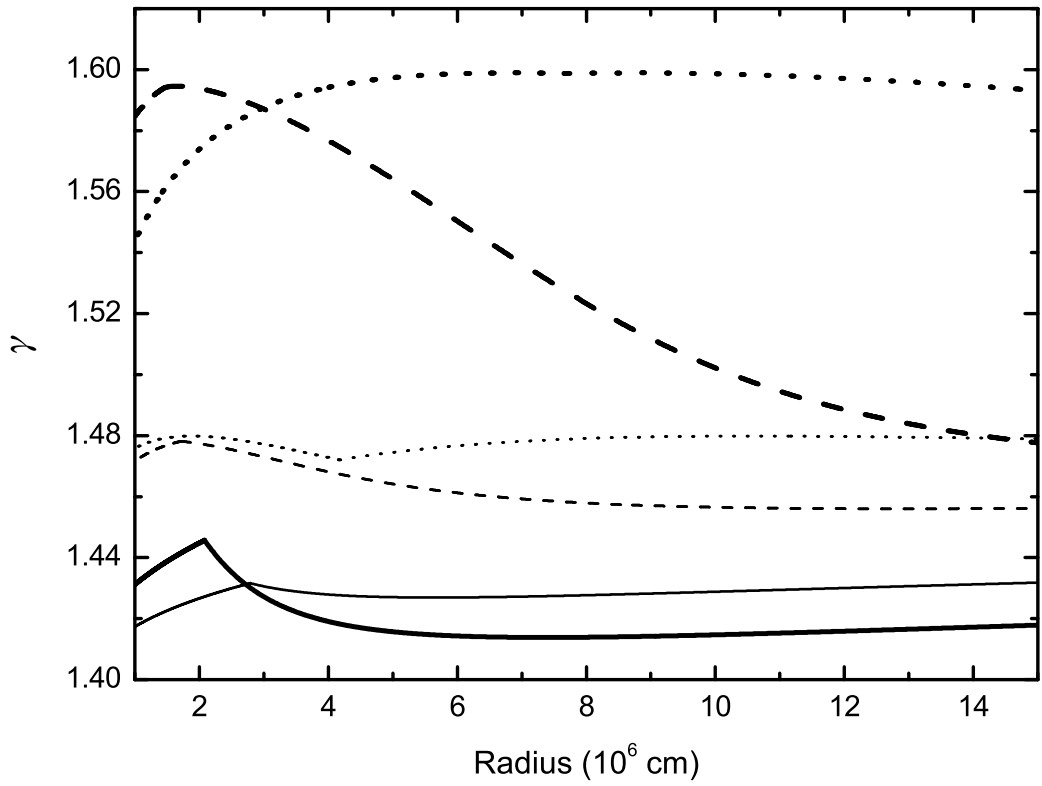}} \caption{The
``equivalent" adiabatic index $\gamma$ of the disk as a function of
radius $r$ in the simple model. The meanings of different lines are
the same as those in Fig. 2.}
\end{figure}

\newpage
\begin{figure}
\resizebox{\hsize}{!} {\includegraphics{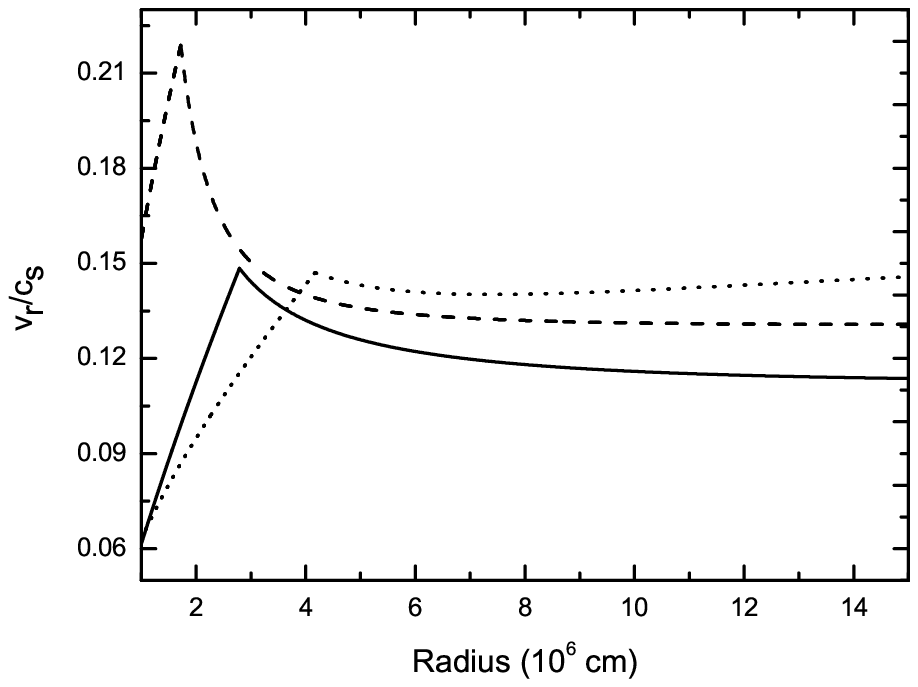}
\includegraphics{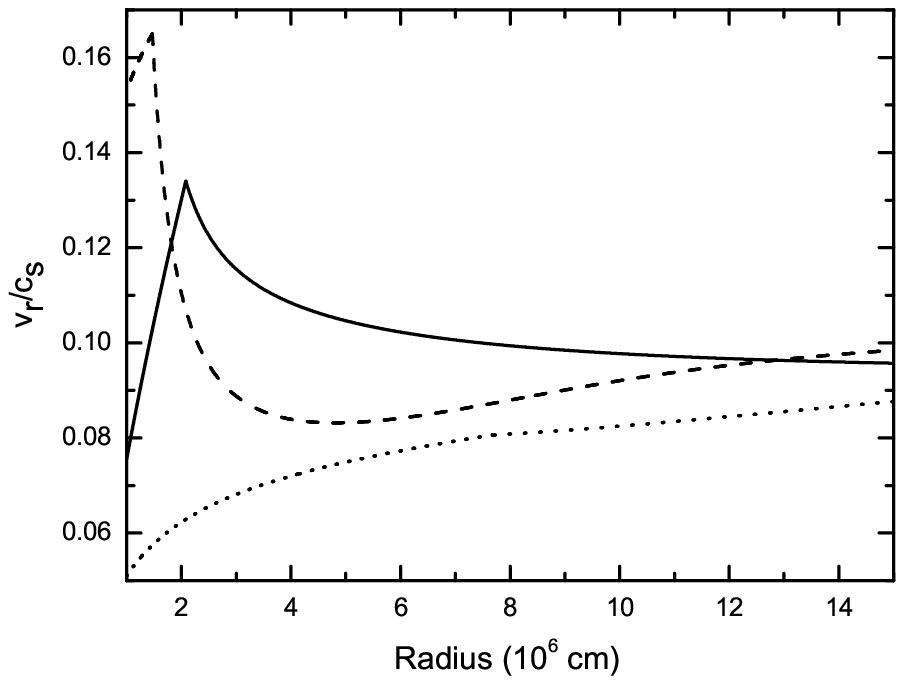}}
\caption{The ratio of the radial velocity $v_{r}$ and local speed of
sound $c_{s}$ as a function of $r$. (a) {\em Left panel}:
$M=1.4M_{\odot}$, $Y_{e}=1.0$; (b) {\em Right panel}:
$M=1.4M_{\odot}$, $Y_{e}=1/9$. The accretion rate:
$\dot{m}_{d}=0.01M_{\odot}\,{\rm s}^{-1}$ ({\em solid line}),
$\dot{m}_{d}=0.1M_{\odot}\,{\rm s}^{-1}$ ({\em dashed line}), and
$\dot{m}_{d}=1.0M_{\odot}\,{\rm s}^{-1}$ ({\em dotted line})}
\end{figure}

\newpage
\begin{figure}
\resizebox{\hsize}{!} {\includegraphics{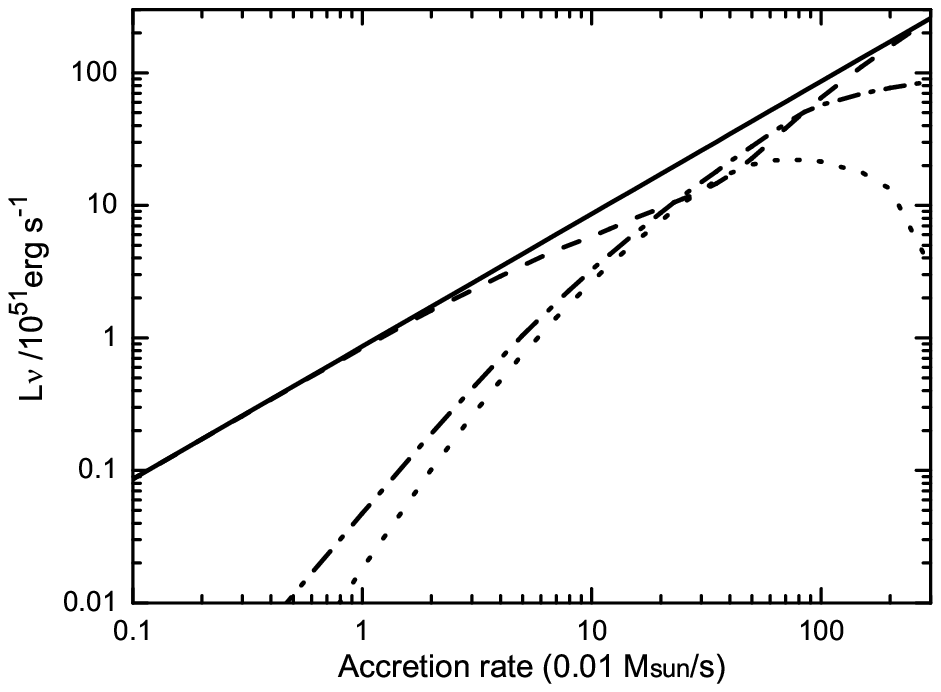}
\includegraphics{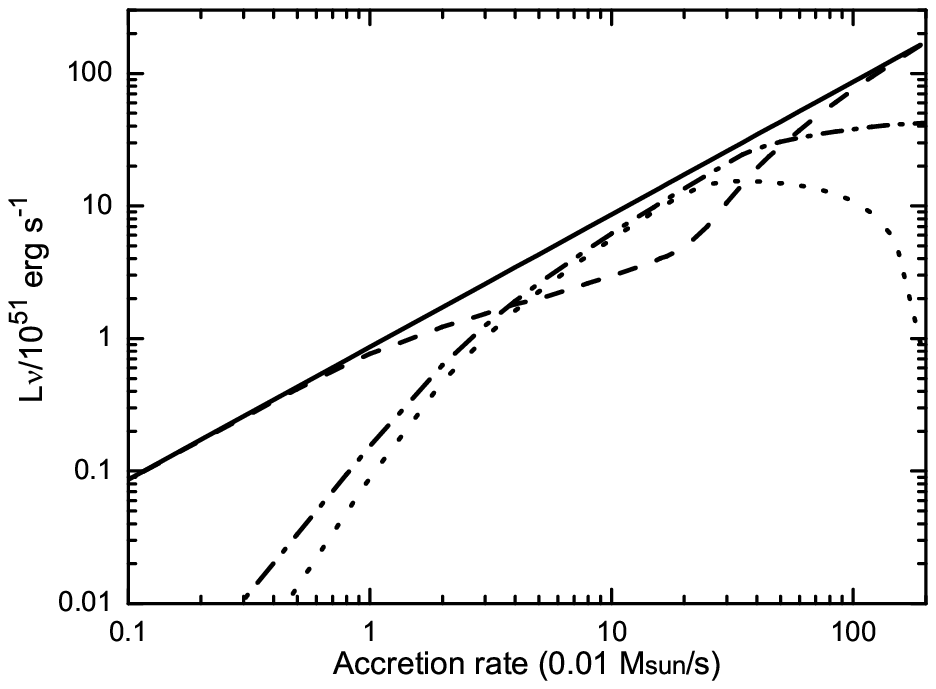}}
\caption{Neutrino luminosity from the disk around a neutron star in
the simple model. $L_\nu$ is in units of $10^{51}$ergs\,s$^{-1}$.
(a) {\em Left panel}: $M=1.4M_{\odot}$ and $Y_{e}=1.0$. (b) {\em
Right panel}: $M=1.4M_{\odot}$ and $Y_{e}=1/9$. The solid line
corresponds to the maximum energy release rate of the disk around a
neutron star, the dashed line to the neutrino luminosity from the
inner disk, the dotted line to the neutrino luminosity from the
outer disk, and the the dash-dotted line to the neutrino luminosity
from a black hole disk.}
\end{figure}
\newpage
\begin{figure}
\resizebox{\hsize}{!} {\includegraphics{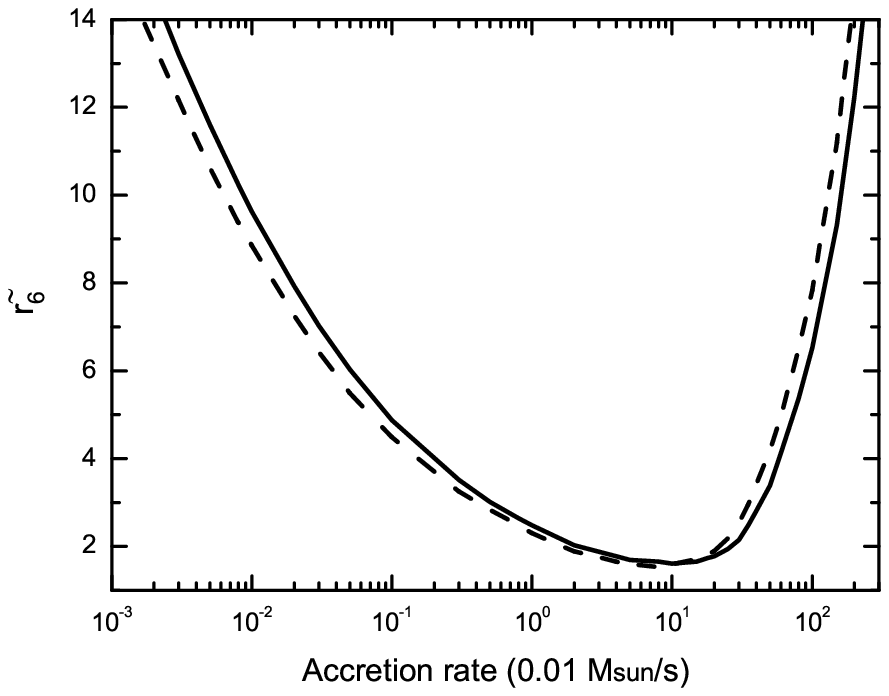}
\includegraphics{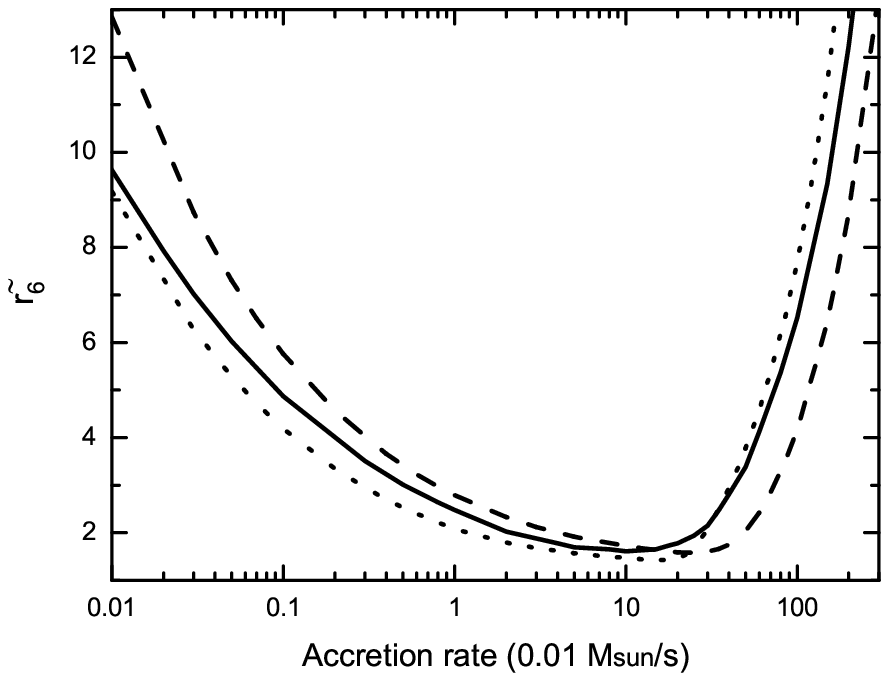}}
\caption{(a) {\em Left panel}: the radius $\tilde{r}_{6}$ between
the inner and the outer disks in the elaborate model with
$M=1.4M_{\odot}$ ({\em solid line}), and $M=2.0M_{\odot}$ ({\em
dashed line}). (b) {\em Right panel}: comparison of $\tilde{r}_{6}$
in different models with $M=1.4M_{\odot}$: (1) the elaborate model
({\em solid line}), (2) the simple model with $Y_{e}=1$ ({\em dashed
line}), and (3) the simple model with $Y_{e}=1/9$ ({\em dotted
line}).}
\end{figure}

\newpage
\begin{figure}
\resizebox{\hsize}{!} {\includegraphics{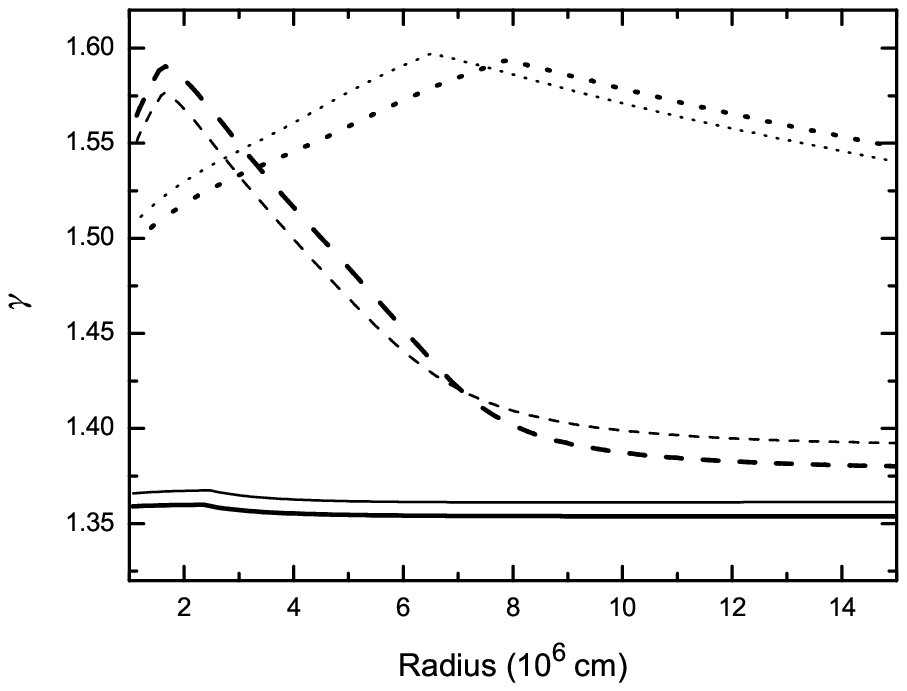}
\includegraphics{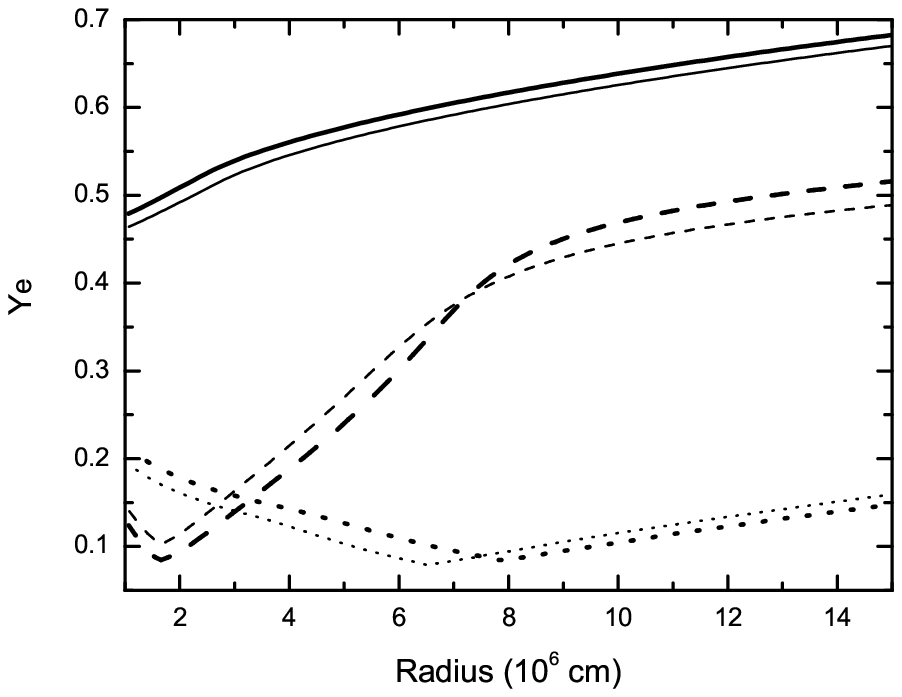}}
\caption{(a) {\em Left panel}: the ``equivalent" adiabatic index
$\gamma$ in the elaborate model. (b) {\em Right panel}: the electron
fraction $Y_{e}$ in the elaborate model as a function of radius. The
profiles are shown for three values of the accretion rate and two
values of the central neutron star mass: (1) $M=1.4M_{\odot}$ and
$\dot{M}=0.01M_{\odot}\,{\rm s}^{-1}$ ({\em thin solid line}), (2)
$M=1.4M_{\odot}$ and $\dot{M}=0.1M_{\odot}\,{\rm s}^{-1}$ ({\em thin
dashed line}), (3) $M=1.4M_{\odot}$ and $\dot{M}=1.0M_{\odot}\,{\rm
s}^{-1}$ ({\em thin dotted line}), (4) $M=2.0M_{\odot}$ and
$\dot{M}=0.01M_{\odot}\,{\rm s}^{-1}$ ({\em thick solid line}), (5)
$M=2.0M_{\odot}$ and $\dot{M}=0.1M_{\odot}\,{\rm s}^{-1}$ ({\em
thick dashed line}), and (6) $M=2.0M_{\odot}$ and
$\dot{M}=1.0M_{\odot}\,{\rm s}^{-1}$ ({\em thick dotted line}).}
\end{figure}

\newpage
\begin{figure}
\resizebox{\hsize}{!} {\includegraphics{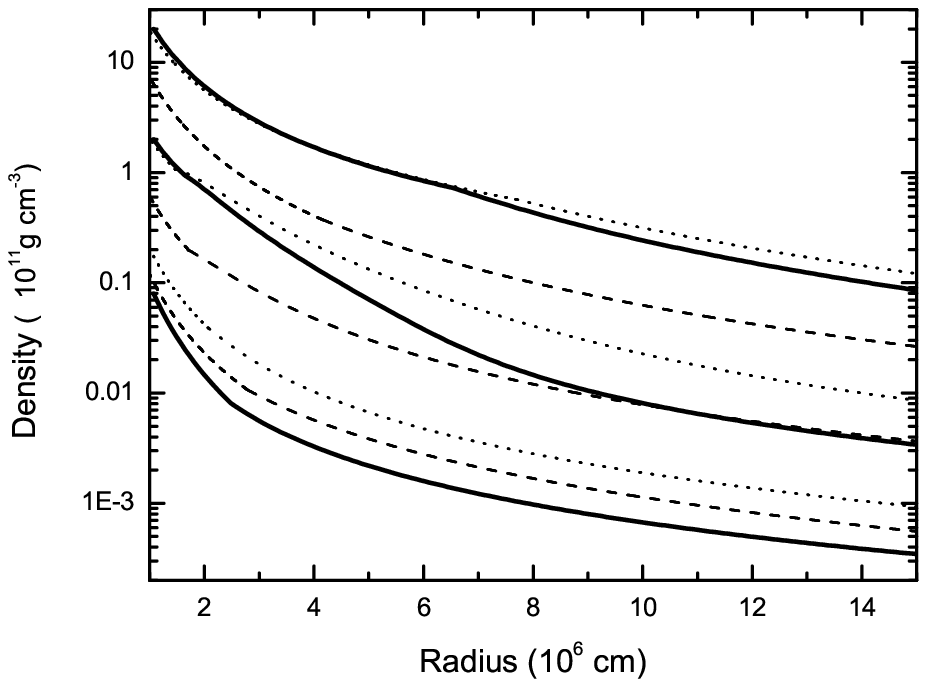}
\includegraphics{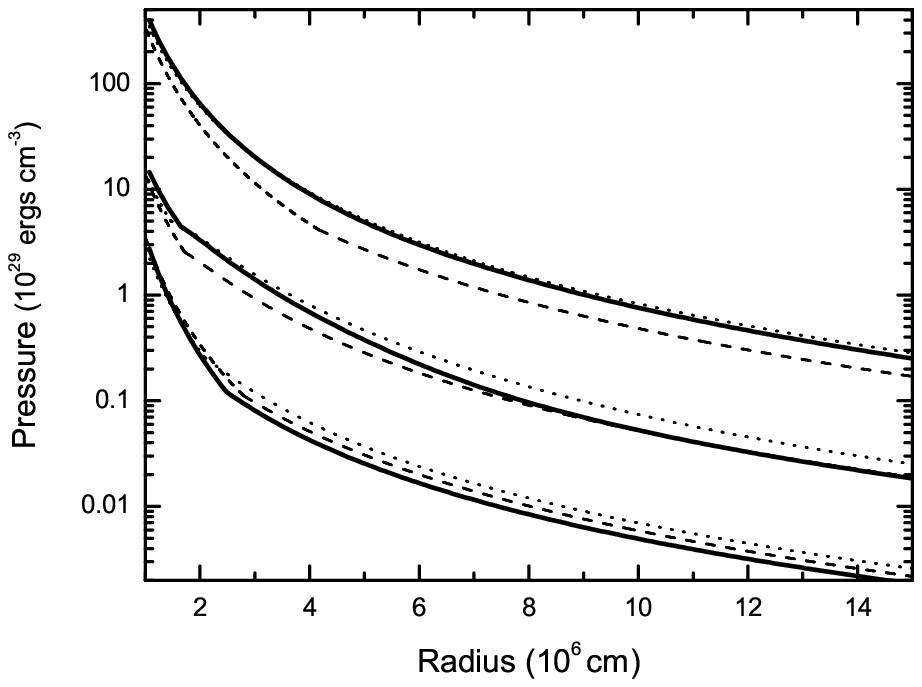}}
\\\resizebox{\hsize}{!} {\includegraphics{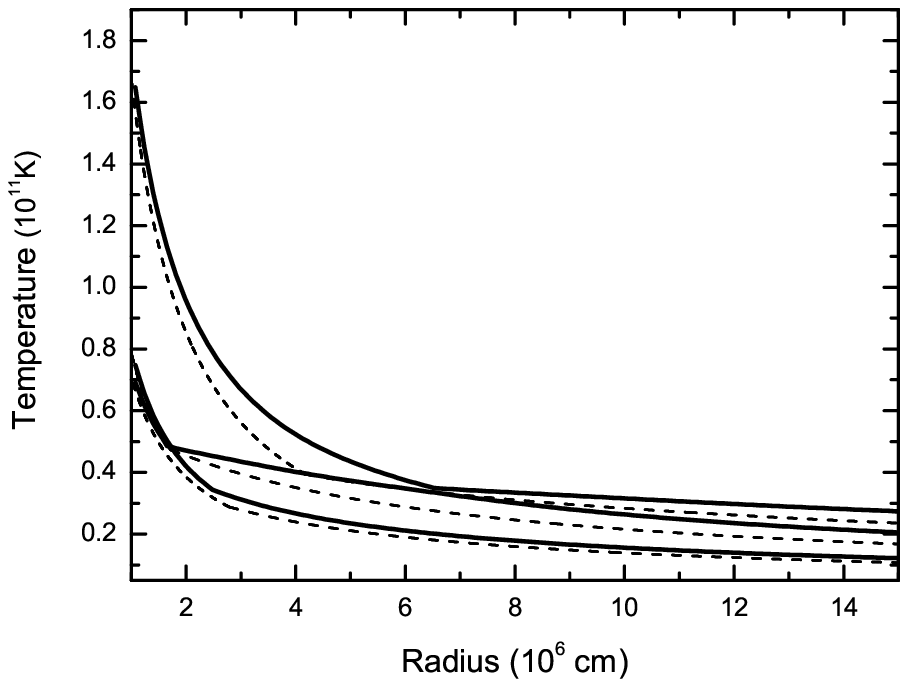}
\includegraphics{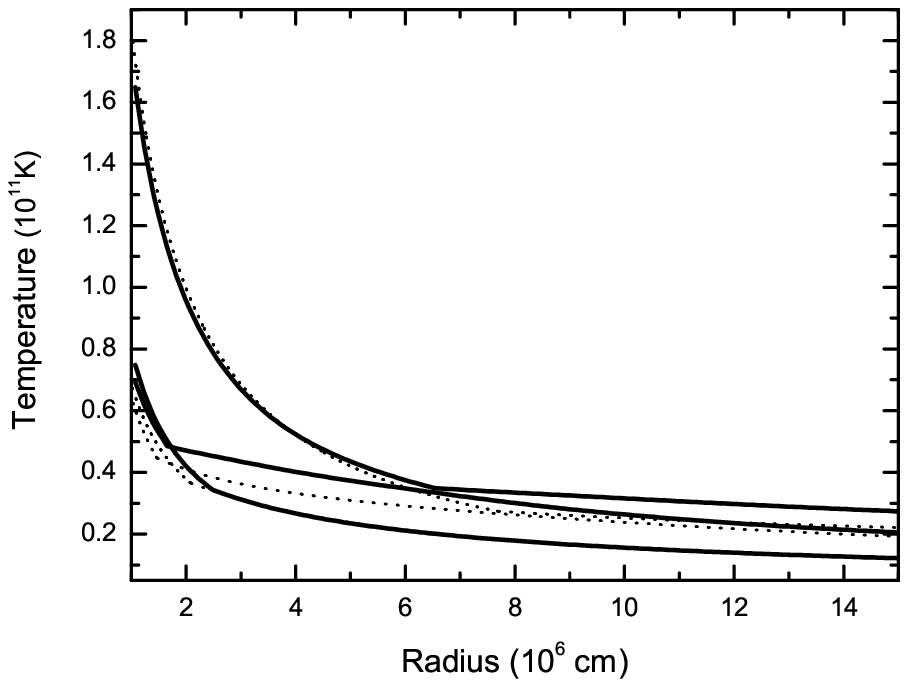}}
\caption{The density (in units of $10^{11}$g cm$^{-3}$), pressure
(in units of $10^{11}$ ergs cm$^{-3}$) and temperature (in units of
$10^{11}$K) of the entire disk in two models for $M=1.4M_{\odot}$.
The profiles include three groups of lines and each group also
includes three lines, which are shown for three values of the
accretion rate $\dot{M}=0.01M_{\odot}\,{\rm s}^{-1}$,
$0.1M_{\odot}\,{\rm s}^{-1}$ and $1.0M_{\odot}\,{\rm s}^{-1}$ from
bottom to top in these figures. The solution in the simple model
with $Y_{e}=1$ is shown by the thin dashed line, and $Y_{e}=1/9$ by
the thin dotted line. The solution in the elaborate model is shown
by the thick solid line. }
\end{figure}

\newpage
\begin{figure}
\resizebox{\hsize}{!}
{\includegraphics{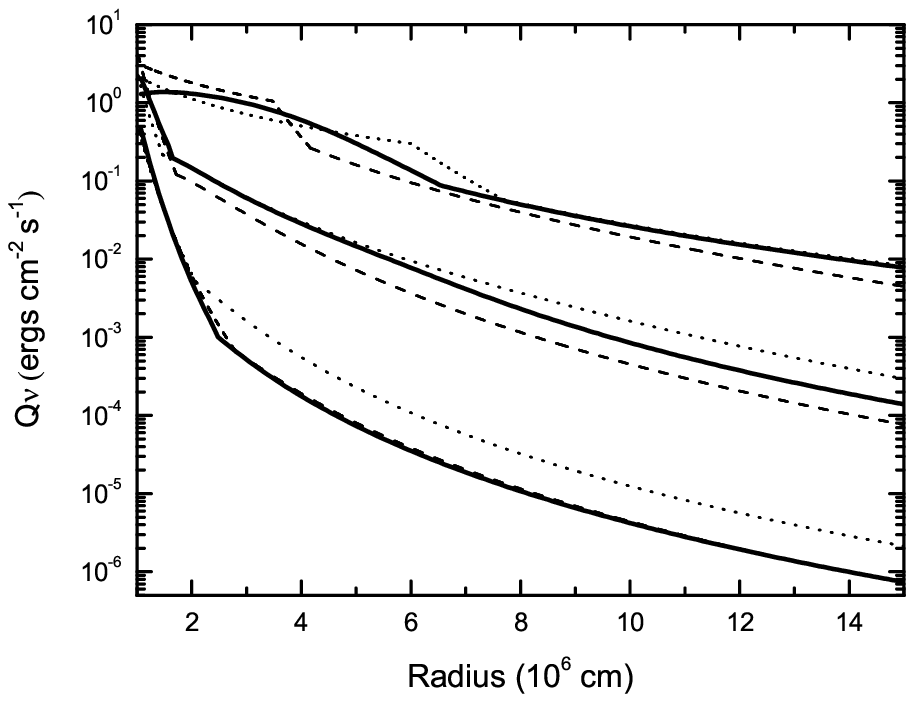}\\\includegraphics{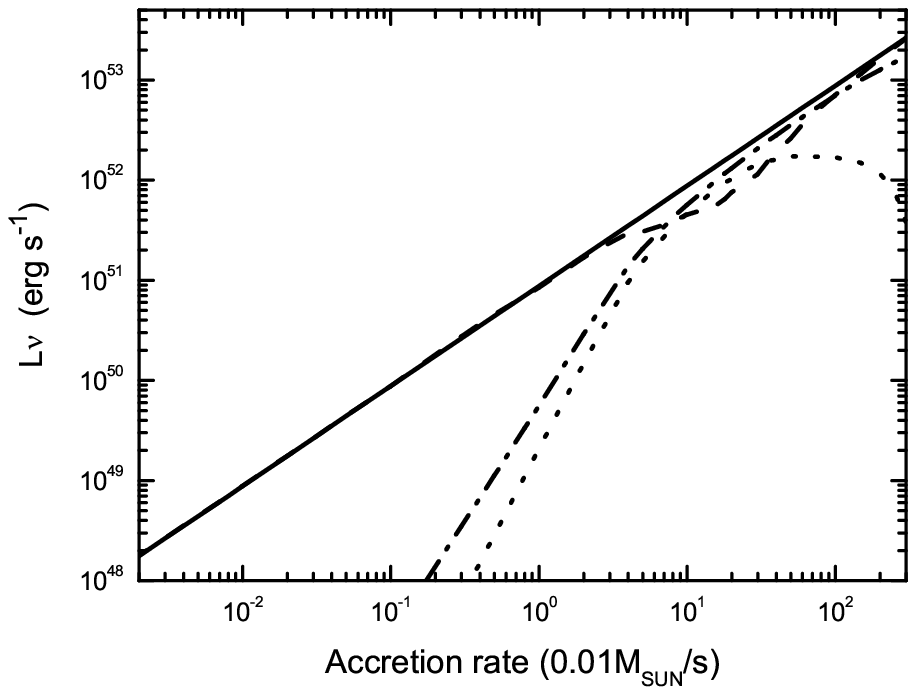}}
\caption{(a) {\em Left panel}: the neutrino luminosity per unit area
(in units of $10^{39}\,{\rm ergs}\,{\rm cm}^{-2}\,{\rm s}^{-1}$) in
both the simple model and the elaborate model. The meanings of
different lines are the same as those in Fig. 8. (b) {\em Right
panel}: the neutrino luminosity from the disk in the elaborate
model. The meanings of different lines are the same as those in Fig
.5.}
\end{figure}

\newpage
\begin{figure}
\resizebox{\hsize}{!} {\includegraphics{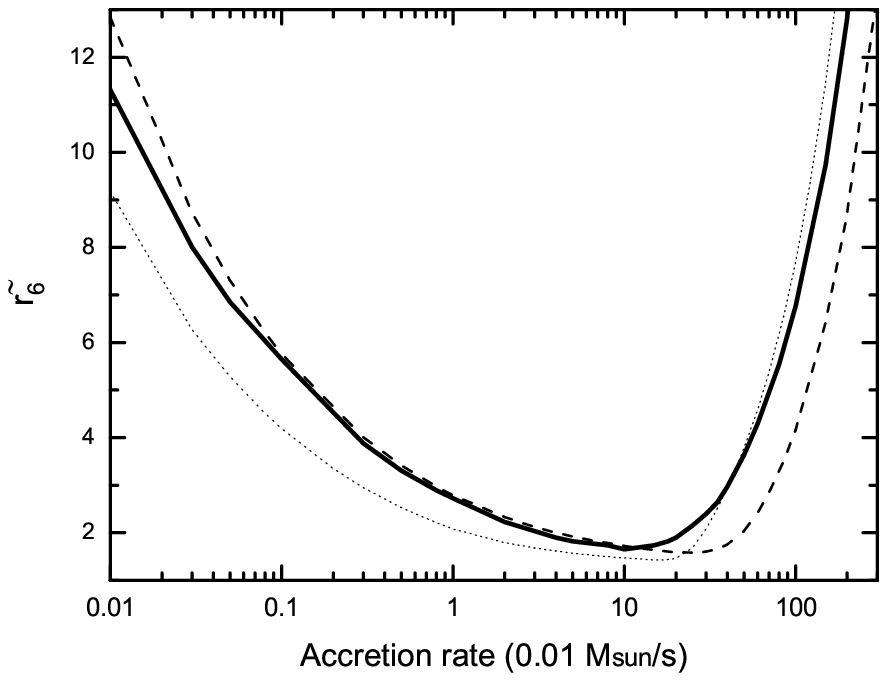}}
\caption{Comparison of $\tilde{r}_{6}$ in three models with
$M=1.4M_{\odot}$: the simple model with $Y_{e}=1$ ({\em dashed
line}), the simple model with $Y_{e}=1/9$ ({\em dotted line}), and
the third model discussed in \S5 ({\em thick solid line}). The
solution of $\tilde{r}_{6}$ in the third model is even more
consistent with the simple model than that of the elaborate model
discussed in \S4.}
\end{figure}


\begin{thebibliography}{}
\bibitem{}Beloborodov, A. M. 2003, ApJ, 588, 931
\bibitem{}Bethe, H. A. 1990, Rev. Mod. Phys., 62, 801
\bibitem{}Brown, G. E., \& Weingartner, J. C. 1994, ApJ, 436, 834
\bibitem{}Chen, W., \& Beloborodov, A. M. 2007, ApJ, 657, 383
\bibitem{}Chevalier, R. A. 1989, ApJ, 346, 847
\bibitem{}Chevalier, R. A. 1996, ApJ, 459, 322
\bibitem{}Dai, Z. G. 2004, ApJ, 606, 1000
\bibitem{}Dai, Z. G., \& Lu, T. 1998a, A\&A, 333, L87
\bibitem{}Dai, Z. G., \& Lu, T. 1998b, Phys. Rev. Lett., 81, 4301
\bibitem{}Dai, Z. G., Wang, X. Y., Wu, X. F., \& Zhang, B. 2006, Science, 311, 1127
\bibitem{}Di Matteo, T., Perna, R., \& Narayan, R. 2002, ApJ, 579, 706
\bibitem{}Fan, Y. Z., \& Xu, D. 2006, MNRAS, 372, L19
\bibitem{}Frank, J., King, A., \& Raine, D. 2002, Accretion Power in Astrophysics
          (Cambridge: Cambridge Univ. Press)
\bibitem{}Gu, W.-M., Liu, T., \& Lu, J.-F. 2006, ApJ, 643, L87
\bibitem{}Janiuk, A., Yuan, Y., Perna, R., \& Di Matteo, T. 2007, ApJ, 664, 1011
\bibitem{}Klu\'zniak, W., \& Ruderman, M. 1998, ApJ, 505, L113
\bibitem{}Klu\'zniak, W., \& Wilson J. R. 1991, ApJ, 372, L87
\bibitem{}Kohri, K., \& Mineshige, S. 2002, ApJ, 577, 311
\bibitem{}Kohri, K., Narayan, R., \& Piran, T. 2005, ApJ, 629, 341
\bibitem{}Lee, W. H., \& Ramirez-Ruiz, E. 2007, arXiv: astro-ph/0701874v3
\bibitem{}Liang, E. W., Zhang, B.-B., \& Zhang, B. 2007, ApJ, 670, 565
\bibitem{}Liu, T., Gu, W. M., Xue, L., \& Lu, J. F. 2007, ApJ, 661, 1025
\bibitem{}Mazzali, P. A. et al. 2006, Nature, 442, 1018
\bibitem{}Medvedev, M. V. 2004, arXiv:astro-ph/0407062
\bibitem{}Medvedev, M. V., \& Narayan, R. 2001, ApJ, 554, 1255
\bibitem{}M\'esz\'aros, P. 2006, Rep. Prog. Phys., 69, 2259
\bibitem{}Nakar, E. 2007, Phys. Rep., 442, 166
\bibitem{}Narayan, R, Igumenshchev, I. V., \& Abramowicz, M. A. 2000, ApJ, 539, 798
\bibitem{}Narayan, R, Mahadevan, R., \& Quataert, E. 1998, in The Theory of Black Hole
          Accretion Discs, ed. M. A. Abramowicz, G. Bjornsson, \& J. E. Pringle
          (Cambridge: Cambridge Univ. Press)
\bibitem{}Narayan, R., Paczynski, B., \& Piran, T. 1992, ApJ, 395, L83
\bibitem{}Narayan, R., Piran, T., \& Kumar, P. 2001, ApJ, 557, 949
\bibitem{}Narayan, R., \& Yi, I. 1994, ApJ, 428, L13
\bibitem{}Paczy\'nski, B., \& Haensel, P. 2005, MNRAS, 362, L4
\bibitem{}Perna, R., Armitage, P. J., \& Zhang, B. 2006, ApJ, 636, L29
\bibitem{}Piran, T. 2004, Rev. Mod. Phys., 76, 1143
\bibitem{}Proga, D., \& Zhang, B. 2006, MNRAS, 370, L61
\bibitem{}Popham, R., \& Narayan, R., 1995, ApJ, 442, 337
\bibitem{}Popham, R., Woosley, S. E, \& Fryer, C. 1999, ApJ, 518, 356
\bibitem{}Shakura, N. I., \& Sunyaev, R. A. 1973, A\&A, 24, 337
\bibitem{}Shapiro, S. L., \& Salpeter, E. E. 1975, ApJ, 198, 671
\bibitem{}Soderberg, A. M. et al. 2006, Nature, 442, 1014
\bibitem{}Spruit, H. C., Matsuda, T., Inoue, M., \& Sawada, K. 1987, MNRAS, 229, 517
\bibitem{}Wang, X. Y., Dai, Z. G., Lu, T., Wei, D. M., \& Huang, Y. F. 2000, A\&A, 357, 543
\bibitem{}Wheeler, J. C., Yi, I., Hoflich, P., \& Wang, L. 2000, ApJ, 537, 810
\bibitem{}Yu, Y. W., \& Dai, Z. G. 2007a, A\&A, 470, 119
\bibitem{}Yu, Y. W., Liu, X. W., \& Dai, Z. G. 2007b, ApJ, 671, 637
\bibitem{}Yuan, Y. F. 2005, Phys. Rev. D, 72, 013007
\bibitem{}Zhang, B. 2007, Chin. J. Astron. Astrophys., 7, 1
\bibitem{}Zhang, B., \& M\'esz\'aros, P. 2001, ApJ, 552, L35
\bibitem{}Zhang, B., \& M\'esz\'aros, P. 2004, IJMPA, 19, 2385

\end{thebibliography}
\end{document}